\documentclass[%
reprint,
superscriptaddress,
amsmath,
amssymb,
aps,
pre,
showpacs,
]{revtex4-1}

\usepackage[utf8]{inputenc}
\usepackage{graphicx}
\usepackage{dcolumn}
\usepackage{bm}
\usepackage{tikz}
\usepackage{enumitem}

\usepackage{pgfplots}
\pgfplotsset{compat=1.3}
\usetikzlibrary{plotmarks}
\usepackage{hyperref}
\renewcommand{\th}[1]{\textcolor{black}{#1}}
\begin{document}
\title{Filamentous phages as building blocks for reconfigurable and hierarchical self-assembly}

\author{Thomas Gibaud}
 \email{thomas.gibaud@ens-lyon.fr}
 \affiliation{Univ Lyon, Ens de Lyon, Univ Claude Bernard, CNRS, Laboratoire de Physique, F-69342 Lyon, France}%

\date{\today}

\begin{abstract}
Filamentous bacteriophages such as fd-like viruses are monodisperse rod-like colloids that have well defined properties: diameter, length, rigidity, charge and chirality. Engineering those viruses leads to a library of colloidal rods which can be used as building blocks for reconfigurable and hierarchical self-assembly. Their condensation in aqueous solution \th{with additive polymers which act as depletants to induce} attraction between the rods leads to a myriad of fluid-like micronic structures ranging from isotropic/nematic droplets, colloid membranes, achiral membrane seeds, twisted ribbons, $\pi$-wall, pores, colloidal skyrmions, Möbius anchors, scallop membranes to membrane rafts. Those structures and the way they shape shift not only shed light on the role of entropy, chiral frustration and topology in soft matter but it also mimics many structures encountered in different fields of science. \th{On one hand}, filamentous phages being an \th{experimental realization} of colloidal hard rods, their condensation mediated by depletion interactions constitutes a blueprint for self-assembly of rod-like particles and provides fundamental foundation for bio- or material oriented applications. \th{On the other hand, the chiral properties of the viruses restrict the generalities of some results but vastly broaden the self-assembly possibilities}.
\end{abstract}



\maketitle

\tableofcontents

\section{Introduction}

Self-assembly is the phenomenon in which a collection of particles spontaneously arranges into mesoscopic structures \cite{whitesides2002, philp1996, rogers2016, frenkel2015, cademartiri2015}. The complexity of the mesoscopic structure is encoded is the building blocks and the physics that drive its condensation. Although the fundamental and driving questions in self-assembly have persisted from the birth of the field up to the present – Can the global behavior of a system be engineered using local rules set by the building blocks properties? How do local rules interlace with physics principles and determine global behavior? – the complexity has drastically increased over the years.

Complexity is in part driven by the properties of the building blocks. In this respect colloidal building blocks have a peculiar flavor. Contrary to polymers, copolymers, surfactants or molecular liquid crystals \cite{majewski2016, drummond1999}, colloids can be considered as giant atoms \cite{poon2004, xia2000} where the solvent mediate the interactions. Consequences are many fold. First, it is a convenient experimental system. Since colloids can be quite big, both the shape and the structure of the condensate can be visualized at the colloidal level under a microscope. Second colloids and their condensates may be manipulated using microfluidics and external fields such as optical traps. Third, due to their large size, the colloids dynamics is also quite slow which permits to probe local kinetics and mechanisms leading to their condensation and their restructuration. Finally colloids come with a large toolbox that permits to engineer their shape and interactions. The last decades have seen an explosion of strategies to obtain synthetic and biological colloids. Colloidal may come in various shapes from spheres to anisotropic geometries \cite{glotzer2004, glotzer2007} such as colloidal  polyhedras \cite{kraft2009, manoharan2003},  tetrapods \cite{milliron2004}, dumbbells \cite{johnson2005} and  triangles \cite{malikova2002} or cubes \cite{laine1998, rycenga2008}. Classical techniques to tailor isotropic interactions include tuning the Van der Waals attraction via the Hamaker constant, grafting polymers to the colloid surface for entropic repulsion, screening the surface charge  with salt, the use of polymers to induce depletion interactions \cite{israelachvili2011}, etc. \ldots  Anisotropic interactions are common in proteins \cite{gogelein2008} and may be induced just by the shape of the colloid \cite{van2013} or by functionalizing the particle surface \cite{bianchi2011} with lock and key groups, such as topological patches, DNA oligonucleotides \cite{alivisatos1996, mirkin1996, milam2003, park2008}, protein-based cross-linkers like biotin–avidin or antibody–antigen binding pairs \cite{hiddessen2000}, and metallic patches \cite{zhang2005}. Depending on the strength and topology of the patchy interactions, the bounding of two colloids may lead to super colloids with internal degree of freedoms such articulated bonds for instance obtained in 'lock and key' colloids \citep{sacanna2010} or by grafting DNA onto liquid interfaces of emulsions \cite{feng2013}. Anisotropic interactions may also come from the solvent. Small water droplets dispersed in a nematic liquid crystal exhibit short-range repulsion and a long-range dipolar attraction which lead to the formation of anisotropic water droplets chainlike structures \cite{poulin1997}. Interactions that can be triggered externally are essential knobs to study reconfigurable self-assembly and sequential or layer-by-layer self-assembly. They allow to navigate in the phase diagram of colloidal dispersions in a continuous way, providing reversible pathway to induce transition between different structures. Divers strategies are adopted to build interactions that respond to temperature (DNA coated colloids \cite{geerts2010, dreyfus2009}, proteins \cite{gibaud2011}), magnetic fields \cite{lowen2001, lin2001, puntes2001}, or electric fields \cite{leunissen2009, liu2014}.

Complexity also comes from the interplay between thermodynamics and kinetics. Self-assembly is a stochastic process and the thermal energy $k_BT$ plays a particular role. It enables the particles to diffuse and probe the energy landscape of the dispersion. In principle, according to thermodynamics, the preferred self-assembled structures are the ones that minimize the free energy of the dispersion. However the actual structures obtained may depend on non-equilibrium effects, local fields, kinetic traps, and pathway-dependent ordering. Hard spheres crystallization is a simple case of self assembly directed by entropy that can be hindered by kinetic effect, namely the glass transition \cite{pusey1986}. Assembly strategies have complexified in recent years. In directed assembly, external fields are used as template to order the colloids \cite{grzelczak2010, huang2001, liu2014}. In template-assisted self assembly, a substrate is used to order colloids \cite{xia2003, chan2004}. In reconfigurable self-assembly, tunable interactions permits transition from one state to another \cite{saunders2009, gibaud2012}. In programmable self-assembly, information is added to the colloids to direct their organization \cite{wan2016, park2008, nagpal2002, wan2016, salgado2015, tanaka2006}. Sequential self-assembly is based on colloids with selective interactions and their sequential activation to form material through multistep kinetics \cite{kotov1995, di2013}. Those strategies may lead to simple structures like homogeneous crystals or to hierarchical assembly where the building blocks organization takes place over distinct multiple levels leading to material structuration at length scales much larger than the building blocks \cite{lopes2001, he2008, elemans2003, aggeli2001, gibaud2012, gibaud2012b, nystrom2017}.

In this review paper, we first focus on filamentous phages -- fd-like viruses, as model colloidal rods, section \ref{sec:fil}. We then show that in presence of depletants their condensation leads to a  myriad of self-assembled structures, section \ref{sec:con}. Finally we discuss a road map using an isotropic aqueous suspensions of filamentous phages and depletion to rationalize the phages hierarchical and reconfigurable self-assembly upon variations of attraction via the depletion interaction, chirality and rods composition, section \ref{sec:conclusion}.

\section{Filamentous phages as versatile building blocks for self assembly}
\label{sec:fil}

A bacteriophage or phage is a virus that infects and replicates within a bacterium. They were discovered by Twort \cite{twort1915} and  d’H\'erelle \cite{herelle1917} in the early 20$^{th}$ century. Bacteriophages are among the most common and diverse entities in the biosphere \cite{mc2007} and are widely distributed in locations populated by bacterial hosts, such as soil or the intestines of animals or sea water. In the latest, up to $\sim$ 10$^8$ virions per milliliter were found in microbial mats at the water surface,\cite{wommack2000}. The impact of phage research in biology is huge: from novel biochemical mechanisms for replication, maintenance and expression of the genetic material and new insights into origins of infectious disease to their use as therapeutic agents \cite{sulakvelidze2001}. Here, we focus on fd-like phages and their use as building block for self-assembly, \cite{dogic2016}.

\subsection{Synthesis} 

\begin{figure}
	\centering
  \includegraphics[width=8.5cm]{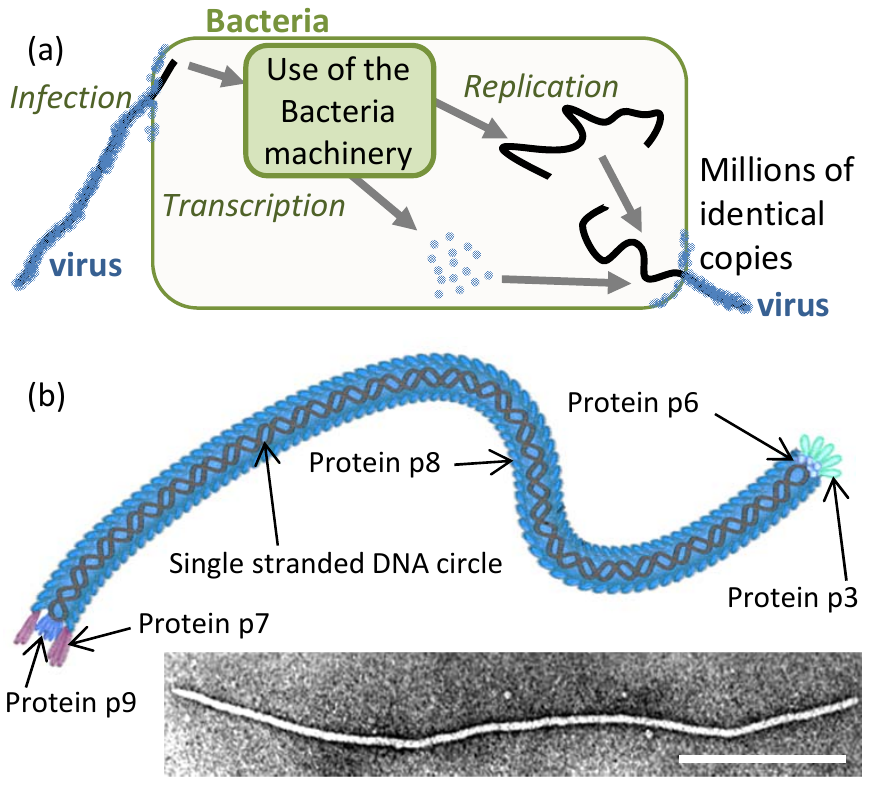}
     \caption{Phage virus like fd-wt use the machinery of E. coli to reproduce itself. (a) Schematic of the reproduction cycle of fd-wt. (b) Schematic, \cite{slonczewski2013} and electron microscopy image of fd-wt virus. Scale bar 200 nm.
     }
    \label{fig:bact}
\end{figure}

The fd-wt virus was originally isolated from sewage \cite{marvin1963}. fd-wt ($M_w=$16.4x10$^6$ g/mol) are identical to one another and composed of a single strand DNA surrounded by a protein layer of about 2700 identical protein p8 subunits. The protein p8 has a molecular mass of 5240 g/mol and accounts for about 99\% of the total protein mass. The rest of the protein mass belongs to the minor coat proteins which are located at the tips of virus \cite{day1988}. At one end of the filament, there are five copies of the protein p9 and p7. At the other end of the phage, there are five copies of p3  and p6. p3 proteins are the first to interact with the E. coli host during infection. p3 is also the last point of contact with the host as a new phage bud from the bacterium, Fig.~\ref{fig:bact}.

\begin{figure}
	\centering
  \includegraphics[width=8.5cm]{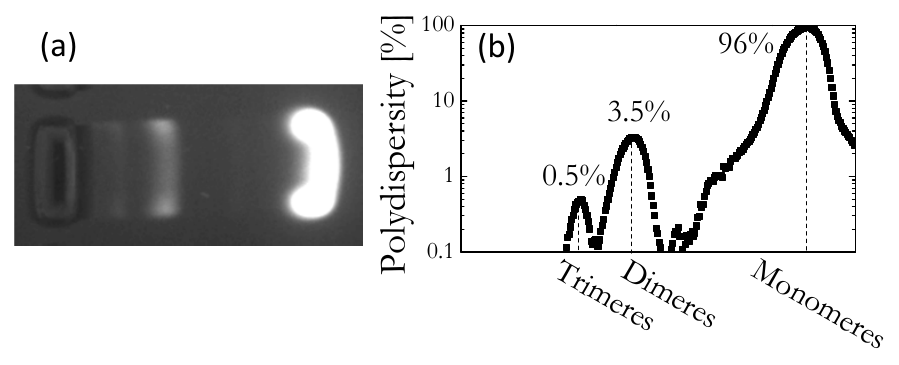}
     \caption{Gel electrophoresis of a typical fd-wt preparation \cite{gibaud2012}. (a) Ethidium Bromide stained gel viewed under UV illumination. The \th{right} bright band consists of 880 nm long fd-wt monomers, while the middle and \th{left} bands contain fd-wt dimers and trimers, respectively. (b) Virus polydispersity is quantified by plotting the normalized intensity profile of the gel, \cite{gibaud2012}.
	}
    \label{fig:gel}
\end{figure}

Bacteriophage viruses were named based on their observed ability to lyse bacterial cells (in greek, ‘bacteria eaters’). However not all phages lyse bacteria. In particular, fd-wt use a lysogenic cycle. Lysogeny is characterized by the integration of the bacteriophage nucleic acids into the host bacterium's genome or formations of a circular replicon in the bacterial cytoplasm. In this condition, the bacterium continues to live and reproduces normally. The genetic material of the bacteriophage is transmitted to daughter cells at each subsequent cell division. Once infected the cell and its descendants are thus turn into a virus manufacture.

fd-wt are grown using standard biological techniques \cite{sambrook1989, unwin2015}. In short, an overnight starter culture taken from a single colony of the bacteria ER2738 is incubated \th{for 12h} at 37$^{\circ}$C and shacked at 250 rpm in 5 mL of sterile 2xYT (yeast extract tryptone) growth medium. 200 $\mu$L from the resulting overnight E. coli culture is then grown in 5 mL of a fresh growth medium until it reaches an optical density $OD=$0.5 at 600 nm measured with a UV-Vis spectrophotometer. The sample is then inoculated with 10 $\mu$L fd-phage stock at approximately 1 mg/mL. The suspension is incubated and shacked 30 minutes then transferred to a 250 mL conical flask with 30 mL of broth media for 2 hours and finally transferred in a 2 L flask with 500 mL of growth media and grown until $OD=$1. From the 500 mL growth cycle, E. coli cells and debris are removed by centrifuging the cultures twice at 8300g for 15 minutes, harvesting the supernatant each time. A precipitant solution is added (146.1 g/L NaCl and 200 g/L \th{Polyethylene glycol (PEG) of average molar mass 8000 g/mol}) in a ratio of 3 parts precipitant to 10 parts supernatant. After refrigeration for at least an hour, the supernatant is centrifuged as before, and the clear supernatant is removed, leaving the precipitated pellet of viruses. Viral pellets are resuspended in 10 mL of sterile phosphate buffered saline (PBS) solution. Ultracentrifuge (1h at 90000 rpm) is then used to exchange the buffer and concentrate the viruses. The virus concentration is determined using absorption spectroscopy. The optical density of fd-wt at 269 nm for 1 mg/ml solution in a 1 cm cuvette is $OD=$3.84 \cite{baus2012}. This procedure yields a virus stock solution with some multimers, such as dimers that have a contour length that is twice that of fd-wt, Fig.~\ref{fig:gel}. To select the monomers, the stock solution is fractionated. Samples are concentrated to reach the isotropic-nematic phase coexistence so that 20\% of the sample is nematic and the rest is isotropic. Longer rods preferentially dissolve in the nematic phase \cite{lekkerkerker1984}. The isotropic fractions is isolated and used as a stock of monodisperse viruses. \th{Such a preparation with 500 mL of growth media yields approximately 200 mg of viruses.} For the self-assembly experiments, the viruses are dispersed in a buffer that contains 100 mM/mL NaCl and 20 mM/mL Tris at pH = 8.05. 

\th{For the study presented in the review, in addition to fd-wt, the filamentous phages fd-y21m and m13KO7 are also used. Their synthesis follow the same protocol as for fd-wt and their unique properties are details in Tab.~\ref{tab:fd}}

\begin{table}
\begin{center}
\begin{tabular}{ c c c c } 
  &fd-wt & fd-y21m & M13KO7 \\ 
 \hline
 $D$ (nm) & 6.6 & 6.6 & 6.6 \\ 
 $L$ ($\mu$m) & 0.88 & 0.88 & 1.2 \\ 
 $L_p$ ($\mu$m) & 2.8 & 9.9 & 2.8  \\
 $C$ ($e^{-}$/nm) & 10 & 10 & 7\\ 
 Chirality & right & left & right\\ 
    \label{tab:fd}
 \end{tabular}
\caption{Properties of the filamentous phages fd, fd-y21m and M13KO7: diameter $D$, Contour length $L$,  Persistence length $L_p$, Charge density at pH=8.05 $C$ and chirality \cite{purdy2004,barry2009,sharma2014,zimmermann1986}.}
\end{center}
\end{table}

\subsection{A model system?}

\begin{figure}
	\centering
  \includegraphics[width=8.5cm]{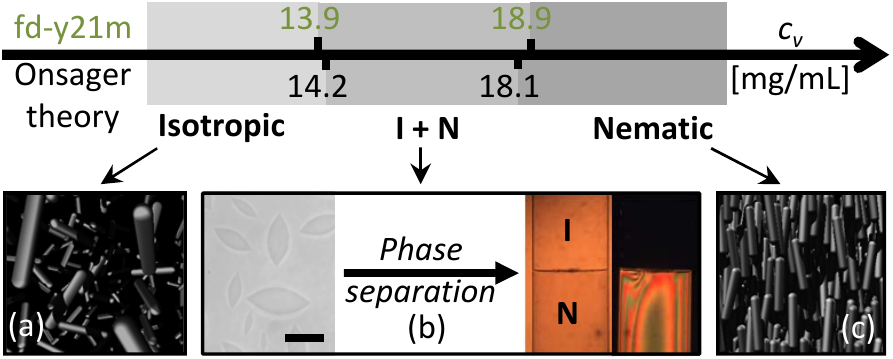}
     \caption{Comparison between the phase diagram of fd-y21m and the Onsager prediction \cite{barry2009}. (a) schematic of the isotropic phase,  (b) I-N phase separation process starting with tactoids and eventually leading to two homegenous isotropic and nematic phases separated by a single interface and (c) a schematic of the nematic phase. Due to their higher flexibility, the coexistence region for fd-wt is narrower and shifted toward higher concentrations: $c_{Iso} =$ 19.8 and $c_{Nem} =$ 22.6 mg/mL. 
	}
    \label{fig:in}
\end{figure}

There are unique advantages of this particular system. First, fd-wt are monodisperse. This eliminates complications related to the polydispersity of rods and facilitates direct quantitative comparison with theory. Second, fd-wt have a diameter of 6.6 nm for a contour length of 880 nm \cite{newman1977, bhattacharjee1992} conferring them a large aspect ratio, $\sim$130 which is similar to the one of spaghetti ($\sim$150). Finally viruses are quite rigid: fd-wt has a persistence length of $\sim$2.8 $\mu$m and the mutant fd-y21m has an even greater persistence length, $\sim$9.9 $\mu$m \cite{barry2009}. Therefore, viruses, and fd-y21m in particular, can be consider as a model liquid crystal system in the framework defined by Onsager. At low concentrations, colloidal rods form an isotropic phase with no direction or orientation order. However as the concentration is increased, the isotropic dispersion becomes metastable or unstable: orientation fluctuations drive concentration gradients which lead to phase separation \th{into a isotropic state in coexistence with a nematic state where the rods have no positional order but tend to point in the same direction which define the nematic director}. During this transition, transient nematic droplets or tactoids  nucleate in an isotropic background and coalesce to minimize the interface between the isotropic and the nematic phases. This leads to the thermodynamically stable state: two homogeneous phases, the isotropic and nematic phase separated by a single interface, Fig.~\ref{fig:in}. Onsager has established that this transition is purely entropic in nature \cite{onsager1949,frenkel1994}. The entropy loss due to the orientation ordering in the nematic phase is over compensated by the increase in translational entropy: the free volume for any one rod increases as the rods align. Moreover, he established that the transition volume fractions for rigid rods with an aspect ratio larger than 75 and repulsive interactions are:  $\phi_{Iso}$ = 3.289$D/L$ for the isotropic phase and $\phi_{Nem}$ = 4.192$D/L$ for nematic phase \cite{onsager1949}. $D$ is the diameter of the rod and $L$ its contour length. \th{For rods with the aspect ratio of fd-y21m, the transition concentrations are $c_{Iso} =$ 14.2 and $c_{Nem} =$ 18.1 mg/mL.} These predictions are remarkably close to the experimental results \cite{barry2009}, Fig.~\ref{fig:in}.

\subsection{A versatile library of colloid rods}
A challenge associated with hierarchical assembly is to control the final macroscopic assemblage by specific modification of relevant microscopic parameters. Thanks to nature diversity, genetic engineering and bio-chemistry, \th{on top of having properties that remains yet unmatched by chemical synthesis}, it is possible create a large library of monodisperse fd-like particles with slight variations in their physical properties, like their contour length, diameter, rigidity or interactions \cite{dogic2016, dogic2014}.

\subsubsection{Chirality}
\begin{figure}
	\centering
  \includegraphics[width=8.5cm]{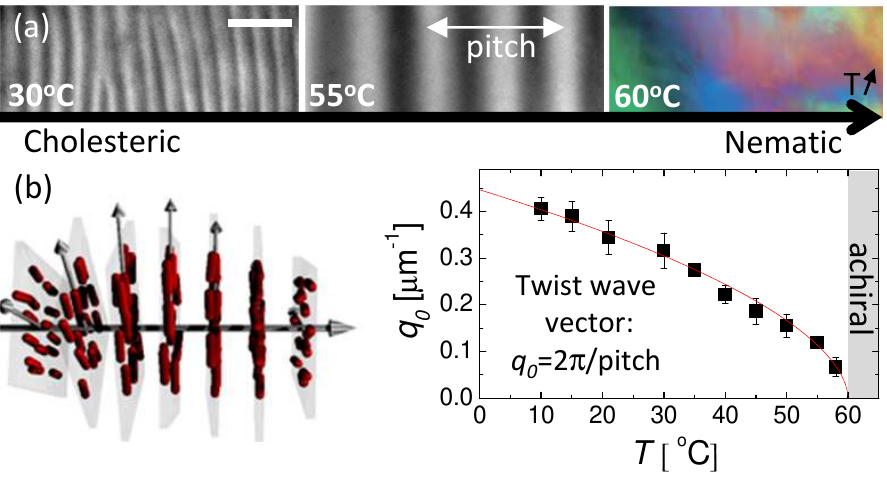}
     \caption{Effect of temperature on the chirality of an aqueous suspensions of fd-wt at $c_v$=100 mg/mL \cite{gibaud2012}. (a) Cross polarized \th{light} microscopy pictures of the nematic and cholesteric phases. Scale bar 30 $\mu$m. (b) Measurements of the twist wave vector $q_0$ of the cholesteric phases as a function of temperature. When $q_0=0$ the cholesteric pitch diverges and the cholesteric phase becomes achiral and nematic. Left: sketch of the cholesteric phase.
	}
    \label{fig:chol}
\end{figure}

fd-wt is chiral an left-handed: in close contact with one another fd-wt tend to twist preferentially clock wise. Therefore fd-wt, at room temperature, form cholesteric phase instead of a nematic \cite{dogic2000, gibaud2012}, Fig. \ref{fig:chol}. \th{The cholestiric phase shows nematic ordering but its director rotates throughout the sample. The axis of this rotation is normal to the director and the distance over which the director rotates by 360$^\circ$ is called the cholesteric pitch}. Moreover, fd-wt chirality is temperature sensitive. Chirality decrease with temperature and eventually vanish at $T=60$ $^{\circ}$C, Fig. \ref{fig:chol}. Understanding the virus chirality and its temperature dependence and its propagation at the macroscopic length scale remains a challenge \cite{dogic2006, grelet2003, tombolato2006, dussi2015, day1988}. Day and Meyer proposed that the cholesteric twist derivates from a ‘cork screw’ shape of the virus due the interplay between its major coat proteins and its DNA backbone \cite{day1988}. 

\subsubsection{DNA backbone}
The contour length of phage virus scales linearly with its genome size. The virus length impact the dynamics of the virus. Maguire \textit{et al.} have shown that rotational diffusion coefficient of rods in the isotropic phase scale linearly with the length \cite{maguire1980}. The virus length also affect the phase diagram. It shifts the location of the isotropic-nematic phase transition toward higher volume fractions and it stabilizes the smectic phase \cite{dogic2001, purdy2004, purdy2007}. So far physicists have used viruses which length range from $\sim$ 0.4 to 1.2 $\mu$m. However, using molecular cloning techniques it is possible to engineer viruses that are as short as 50 nm and as long as 8000 nm \cite{herrmann1980, specthrie1992, marchi2014, brown2015, sattar2015}. 


\subsubsection{Major coat proteins}
The major coat proteins confer to fd-wt a net linear charge density of 10 $e^{-}$/nm at pH=8.05 \cite{zimmerman1986, purdy2004}. It is possible to label the major coat proteins with chemical compounds. This is very convenient to make the virus fluorescent and track their individual dynamics within an assemblage. Coating such as PEG \cite{grelet2016}, SiO$_2$, TiO$_2$, \cite{pouget2013}, PNIPAM \cite{zhang2009}, DNA \cite{ruff2016}, gold \cite{montalvan2014}, carbon nanofiber \cite{szot2016} or fluorescent dyes \cite{zhang2013} obviously increases the diameter of the virus but may also drastically change the interactions between the viruses and therefore the way they self-assemble.

\begin{figure}
	\centering
  \includegraphics[width=8.5cm]{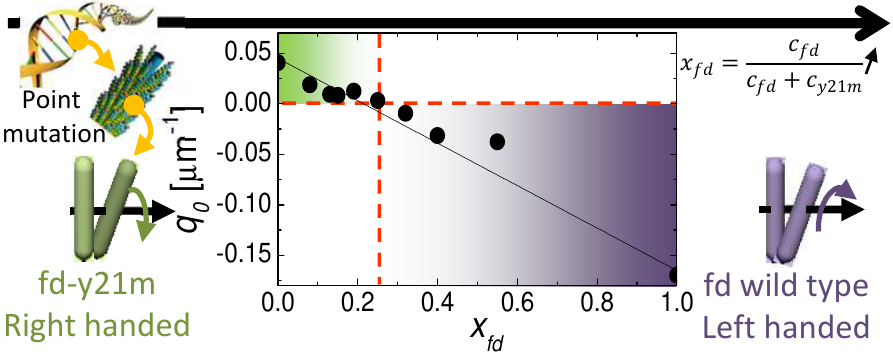}
     \caption{Effect of the ratio of fd-wt to fd-y21m on the chirality of aqueous mixtures of of fd-wt and fd-y21m, $c_v$=55 mg/mL, $T=$ 22 $^{\circ}$C \cite{barry2009}. Measurements of the twist wave vector of the cholesteric phases as a function $x_{fd}=c_{fd}/(c_{fd}+c_{fdy21m})$ where $c_{fd}$ and $c_{fdy21m}$ are the concentration of fd-wt and fd-y21m. Side figures: sketch of the chiral interaction between two fd-wt and the point mutation on the major coat proteins that induce opposite chiral interaction between two fd-y21m.
	}
    \label{fig:chol2}
\end{figure}

Genetic mutation represents another way to act on the major coat proteins \cite{abramov2017}. For example, structural biologists have genetically engineer fd-wt into fd-y21m, a mutant virus in which the 21$^{st}$ amino acid out of the 50 composing the major coat protein is changed from tyrosine to methionine \cite{marvin1994}. fd-y21m is not only stiffer as mention above, but it also makes left-handed cholesteric state as opposed to fd- wt which form right handed cholesteric state \cite{barry2009} and contrary fd-wt, fd-y21m chirality is temperature independent \cite{gibaud2012}. By mixing fd-wt and fd-y21m at a controlled ratio $x_{fd}$, it is possible to design cholesteric phases with the desired the pitch and chirality, Fig. \ref{fig:chol2}. The phase space of all possible mutations of the major coat protein is huge. It could be investigated using phage display technology \cite{rowitch1988} to better understand the impact of the coat protein structure on the coarse-grained properties of the filament \cite{hunter1987}. 

\subsubsection{Cap proteins}
Another attractive feature of filamentous bacteriophages is the presence of cap proteins which are distinct from the major coat proteins, thus enabling selective labeling of the virus end and in particular normal anchoring of the phage, \th{i.e. attaching virus perpendicular to a surface}. On one hand this feature is used to create new materials. For instance using phage display, filamentous phages were organized into smectic layers that are intercalated with layers of end-bound inorganic nanoparticles \cite{whaley2000, lee2002}. On the other hand this feature is also used to design new particles such as star colloids where filamentous phages were pinned to the colloid surface \cite{huang2009} or filamentous ring-like structures by labeling the two ends of the virus with distinct labels that stick to each other \cite{nam2004}. This last example paves the road toward specific and sequential self-assembly.


\section{Condensation of colloidal rods}
\label{sec:con}

\subsection{Colloidal rod and depletion}

\th{The depletion interaction is an effective attraction that arises between large colloidal particles that are suspended in a dilute solution of depletants. Except for excluded volume effects, the depletant and the colloids are not interacting and are considered as hard spheres. Usually the depletant is a polymers much smaller than the colloid. In this configuration, there is a region which surrounds each colloid which is unavailable for the centers of mass of the depletants. Therefore, as two colloids approach each other, the excluded volumes overlap and additional free volume becomes available to the polymers, thus increasing the overall entropy of the mixture \cite{asakura1954}. This results in an effective attractive (depletion) potential between the colloids, whose strength and range can be increased by increasing the polymer concentration and size, respectively \cite{asakura1954}.}

\begin{figure}
	\centering
  \includegraphics[width=8.5cm]{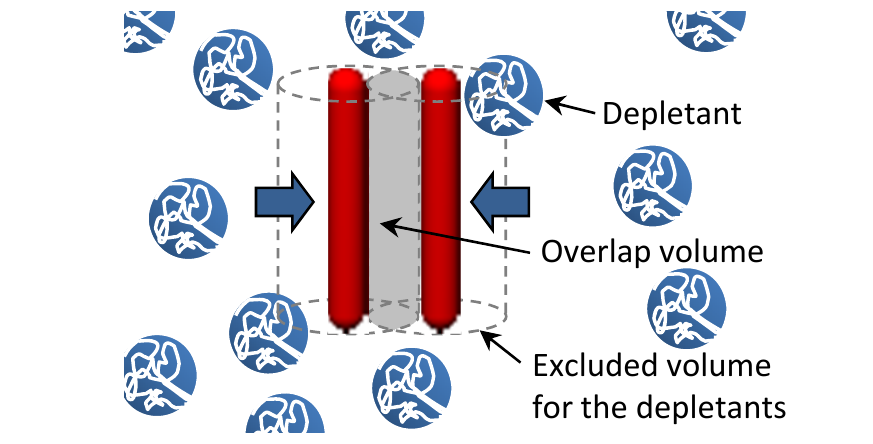}
     \caption{Sketch of the depletion interaction between two colloidal rods.
     }
    \label{fig:dep}
\end{figure}

\begin{figure}
	\centering
  	\includegraphics[width=8.5cm]{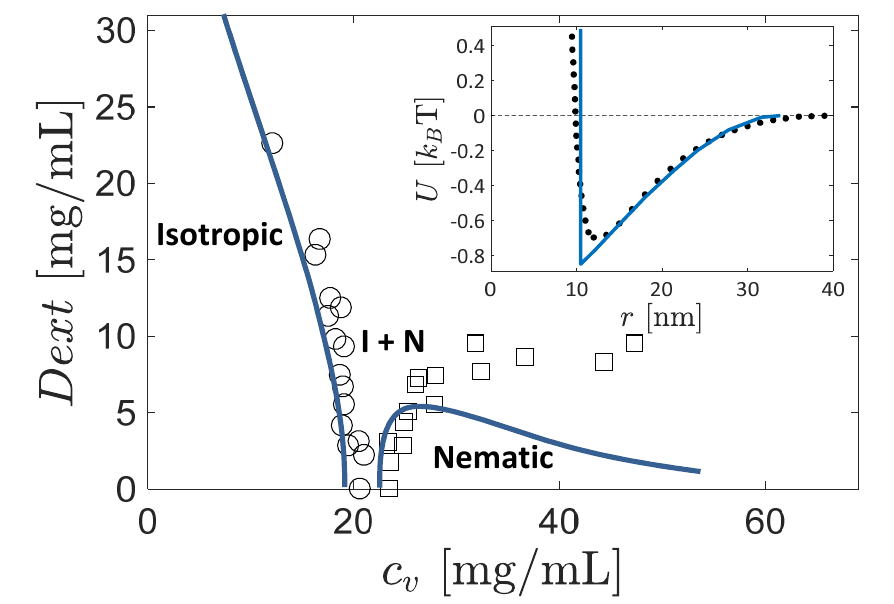}
    \caption{Influence of the depletant dextran ($M_w =$500 000 \th{g/mol}, radius of gyration $R_g=$17.6nm, concentration $Dext$) on an aqueous dispersion of fd-wt at a concentration $c_v$ \cite{dogic2004}. Circles delimit the isotropic state from the I-N coexistence region. Squares delimit the I-N coexistence region from the nematic state. The blue line corresponds to the predictions from the second virial theory with attraction (SVTA). Inset: interaction potential between two viruses separated by a distance $r$ and oriented at 90$^\circ$. The interaction potential is a sum of the electrostatic repulsion and the depletion interaction.  Dot are the results of computer simulation. The line correspond to the approximate intermolecular potential between rods with an effective hard core diameter $D_{eff}=$ 10.5 nm. The attractive part of the potential is modeled by Asakura-Oosawa penetrable spheres whose effective radius and concentration best fits the potential obtained through computer simulation. SVTA use this approximated potential as input. 
	}
    \label{fig:pd}
\end{figure}

\th{The depletion principle can be transposed to mixtures of viruses and polymer depletants. This has been tested under relatively high salt content so that hard-core repulsive interactions dominate and using different polymers like Dextran or Polyethylene glycol which size is always smaller than the rod length $L$ but can be greater than its diameter $D$ \cite{dogic2004, yang2012}. Contrary to colloids/polymer mixtures, the depletion interaction becomes anisotropic with rod-like colloids. It tends to align the viruses \cite{matsuyama2001} so that the overlap volume is maximized Fig.~\ref{fig:dep}.} The obvious consequence in the introduction of polymers in suspensions of phages is that it tends to shift the isotropic boundary to lower volume fractions. Due to the level rule, the nematic state is consequently shifted to higher volume fractions. Using fd-wt and dextran mixtures this behavior is quantitatively confirmed and modeled \cite{dogic2004}, Fig. \ref{fig:pd}. Depletion is an ideal tool to promote entropic condensation.

\subsubsection{Nematic droplets}
\th{Nematic droplets rather than being spherical display a spindle shape \cite{ichinose2004, davidson2010, kaznacheev2002}. This shape is due to the interplay between the interfacial tension and the splay and bend elastic constants of the inner nematic phase \cite{onsager1949}}. Tuning the morphology and order within the droplets represent a corner stone for applications such as light modulators or more generally as photonic materials \cite{tortora2011, jeong2014, bunning2000, joannopoulos2011}.

\begin{figure}
	\centering
  	\includegraphics[width=8.5cm]{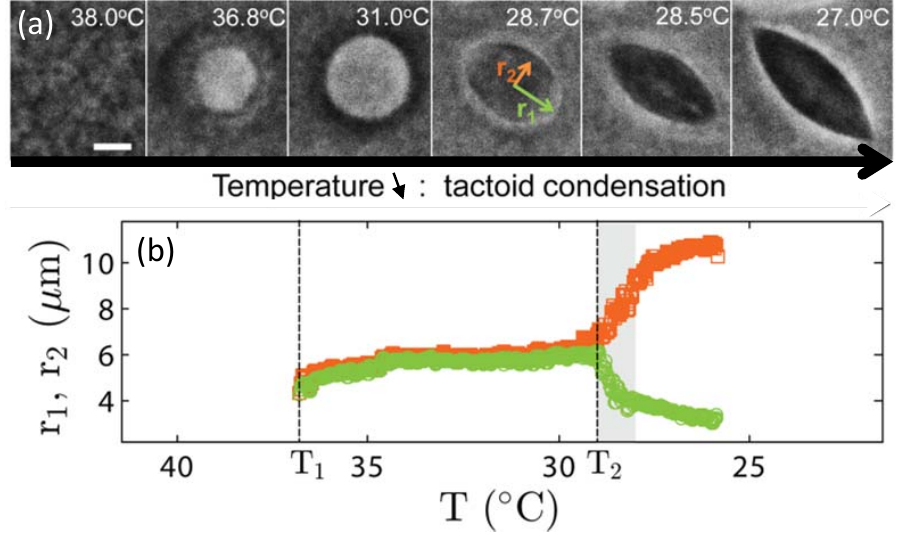}
    \caption{Condensation of nematic droplets in dispersions of colloidal rods with thermo-sensitive depletants \cite{modlinska2015}. (a) Phase contrast images of a suspension of M13K07 viruses at $c_v=$  1mg/mL mixed with pnipam microgel particles at $c_P=$ 30mg/mL. As temperature is lowered (0.09$^\circ$C/min), the attraction increases and the condensation of nematic droplets is observed. The process is fully reversible when temperature is increased. Scale bar, 5 $\mu$m. (b) Measurements of the long and short semi-axis of the droplet, $r_1$ and $r_2$ as a function of temperature $T$. $T_1$ and $T_2$ are respectively the temperatures that separate the isotropic state, the spherical  \th{dense isotropic droplets in an isotropic background regime and the tactoids in an isotropic background regime}.
     }
    \label{fig:tac}
\end{figure}

Most paths which lead to tactoids formation are kinetically driven. For example, Lettinga \textit{et al.} prepared samples in the I-N coexistence region and used shear to dissolve the tactoids and then study their condensation. To circumvent kinetics issues, Modlińska \textit{et al.} engineered a colloidal system where one can continuously tune the attraction between the rods to condensate tactoids in a reversible and quasi static way starting from an equilibrium isotropic state \cite{modlinska2015}. They replaced the depletants Dextran by thermo–sensitive and non-adsorbing \th{poly(N-isopropylacrylamide)} (pnipam) microgel particles \cite{still2013}. The effective attraction between the rods is then controlled externally by temperature. \th{As temperature decreases, the microgel particle swell which increases both the range and the depth of the attraction \cite{modlinska2015}}. Navigating the phase diagram in a continuous way, Modlińska \textit{et al.} showed that tactoids formation is preceded by the nucleation and growth of dense isotropic spherical droplets within the isotropic background, Fig.~\ref{fig:tac}. This scenario is analogous to the enhanced protein crystallization slot located above the liquid-liquid phase separation suggested by ten Wolde and Frenkel \cite{ten1997}. Just as the critical density fluctuations and in particular fluctuations of high densities behave as a micro reactor to lower the energy barrier for crystal nucleation, the dense isotropic droplets layout the ideal nucleation spot for the nematic phase. 

\subsubsection{Colloidal membranes}

\begin{figure}
	\centering
  \includegraphics[width=8.5cm]{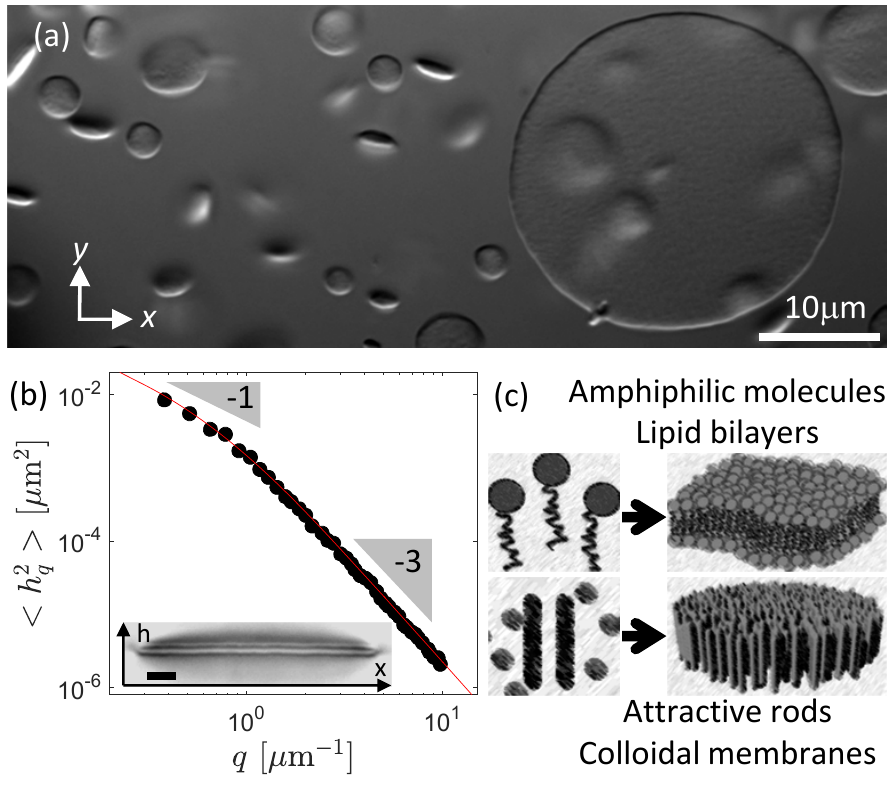}
     \caption{fd-wt colloidal membranes, $T=$ 22$^{\circ}$C, $Dext =$ 45 mg/mL \cite{barry2010}. (a) \th{Differential interference contrast (DIC)} micrograph of colloidal membranes. Scale bar is 10$\mu$m. (b) Fluctuation spectrum resulting from the Fourier analysis of a sequence of uncorrelated membrane conformations. The red line is the best fit to the Helfrich equation, which yields a lateral bending modulus of 135 $k_BT$ and a surface tension of 100 $k_BT$/$\mu$m$^2$. \th{For comparison, the lateral bending modulus and surface tension of lipid membranes are respectively  $\sim 1$ $k_BT$ and $\sim 10^6$ $k_BT$/$\mu$m$^2$}. \cite{gelbart2012, feller1999}. Inset: DIC micrograph of a colloidal membrane edge-on. Scale bar 2 $\mu$m. (c) Sketch of lipid bilayer and colloidal membrane which on a coarse grain level are similar and obey the Helfriech equation.
     }
    \label{fig:mem}
\end{figure}

Fig. \ref{fig:pd} shows that depletion interactions promote rods condensation but it does not reveal any new phases compared to the case without depletant. Barry and Dogic extended this phase diagram to higher dextran concentrations ($M_w=$ 500 000 g/mol) \cite{barry2010} and showed that, starting from an isotropic rods suspension at $c_v=$ 1 to 10 mg/mL, it is possible to assemble a new phase: 2D colloidal membranes composed of a one-rod length thick mono-layer of aligned rods. \th{The membrane diameter is not controlled and varies from a few microns to hundreds of microns, Fig. \ref{fig:mem}}. On a coarse grain level the self-assembled fluid-like and equilibrium monolayers have the same symmetry as lipid bilayers and one can develop many analogies. First, like lipid bilayers, the instantaneous and average projected colloidal membrane area $A$ are proportional, $\langle A-\langle A \rangle \rangle ^2=k_BT \langle A \rangle/\chi $ where the compressibility is $\chi\sim 4500$ $k_BT$/$\mu$m$^2$) \cite{barry2010, nagle1978, needham1990, evans1990, evans1987}. \th{For comparison, the compressibility of lipid membrane is 2 to 3 orders of magnitude higher $\sim 10^7$~$k_BT/\mu$m$^2$\cite{rawicz2000}}. Second, the colloidal membranes viewed in edge-on configurations, exhibit thermal undulations. The Fourier analysis of these fluctuations can be model using the elastic free energy written down by Helfrich, originally developed for lipid bilayers \cite{helfrich1978}. Finally, the stability of colloidal membranes is similarly related to the way lipid bilayers interact \cite{goetz1999, israelachvili1996, lipowsky1993}. Indeed, From $Dext=$45 to 53 mg/mL, colloidal membrane remain isolated from each other: as two membranes approach each other in suspension, protrusion fluctuations lead to an effective repulsive interaction and promote the stability of isolated membranes. At higher dextran concentrations, the depletion interaction becomes sufficiently large to overcome this effective repulsion and colloidal membranes stack on top of each other \cite{barry2010}. 

However, from a microscopic perspective the forces driving the assembly of colloidal membranes and lipid bilayers are very distinct. Colloidal membranes are assembled from micron length hydrophilic rod-like molecules, whereas lipid bilayers are assembled from nanometer amphiphilic lipids. This leads to orders of magnitude difference in their compressibility, lateral bending modulus or lateral tension.\th{Those orders of magnitude differences can in a first approximation be attributed to the size differences of the building blocks. Indeed, the distance $d$ between the constituent particles in the colloidal membrane is $\sim 10$ nm and $\sim 1$ nm is lipid bilayer; assuming that $\chi\sim1/d^2$ \cite{evans1974}, we roughly recover the ratio between the compressibility of colloidal membranes and lipid bilayer. The same holds for the lateral bending modulus which scales as $(D/L)^2$\cite{bermudez2004}}. Colloidal membranes being robust assemblages stable over a wide range of parameters \cite{yang2012}, they represent a unique opportunity to investigate membrane biophysics from an entirely new perspective on length scales where it is possible to visualize and follow under \th{light} microscope the constituent building blocks, the membrane dynamics or reconfigurable processes. 

\begin{figure}
	\centering
  \includegraphics[width=8.5cm]{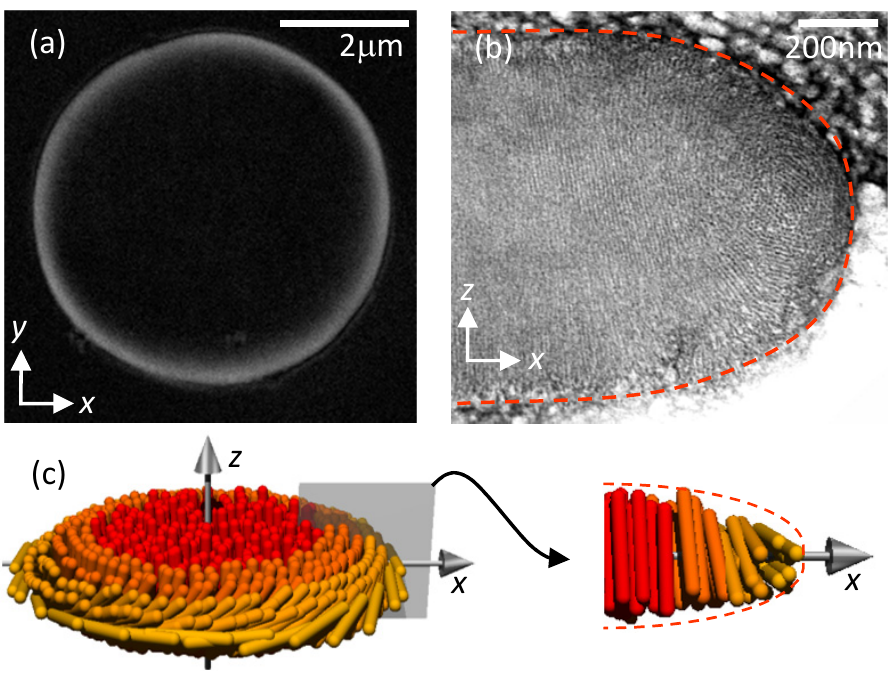}
     \caption{Structure of a colloidal membrane, $T=$ 22$^{\circ}$C, $Dext =$ 45 mg/mL \cite{gibaud2012}. (a) 2D LC-PolScope retardance map of a colloidal membrane.  The bright band associated with the edges indicates local rod tilting. (b) Electron micrograph cross section of a membrane directly visualizing the tilt of the rods and the curved edge profile. (c) Sketch of a colloidal membrane indicating that its edge adopts a curved profile, forcing rods to locally twist.
     }
    \label{fig:edge}
\end{figure}

We first discuss the edge properties of colloidal membranes which are described at a macroscopic level by the interfacial tension $\gamma$ \cite{marchand2011}. For 2D colloidal membranes, $\gamma$ is 1D and is the equivalent of surface tension for 3D objects like emulsion for instance. This a thermodynamic quantity that results from the greater affinity of the colloidal membrane particles to each other than to the particle isolated in the solvent. The net effect is an inward force at the membrane circumference that causes the edge to behave elastically. The control of interfacial tension is manifold. It justifies that colloidal membranes adopt a circular shape. Its control, in analogy with micro-emulsion, could lead to fine tune the size of colloidal membranes.

The edge structure of achiral colloidal membrane is determined using three complementary imaging techniques, namely two-dimensional (2D) and three-dimensional (3D) polarization microscopy and electron microscopy \cite{gibaud2012}, Fig.~\ref{fig:edge}. 2D-LC-PolScope \cite{oldenbourg1995} of a membrane lying normal to the $z$-axis of the microscope produces images in which the intensity of a pixel represents the local retardance and indicates the local tilt of the rods with respect to the $z$-axis. Rods in the bulk of a membrane are aligned along the $z$-axis, and it follows that 2D LC-PolScope images appear black in that region. By contrast, the bright, birefringent ring along the membrane’s periphery reveals local tilting of the rods at the edge. The 3D reconstruction of the membrane structure using electron tomography \cite{mastronarde2005, kremer1996}, shows that the virus tilt by 90$^\circ$, from being normal to the membrane  surface in the bulk to tangential to the edge along the membrane periphery. This behavior is corroborated by 3D-LC-PolScope \cite{oldenbourg2008}. This twist goes with a hemi-toroidal curved edge. The twisted edge makes the membrane a chiral object. For achiral viruses dispersions, the spontaneous twist at the edges is equally likely to be clockwise or anticlockwise \cite{gibaud2012}. For chiral virus suspensions, the edge adopts the chirality of the virus. By comparison with an untilted edge, a curved edge structure lowers the area of the rod-polymer interface, thus reducing interfacial tension, at the cost of increasing the elastic energy due to a twist distortion.

\subsection{Colloidal membranes and chirality}
\subsubsection{Tuning the edge chirality}

\begin{figure}
	\centering
  \includegraphics[width=8.5cm]{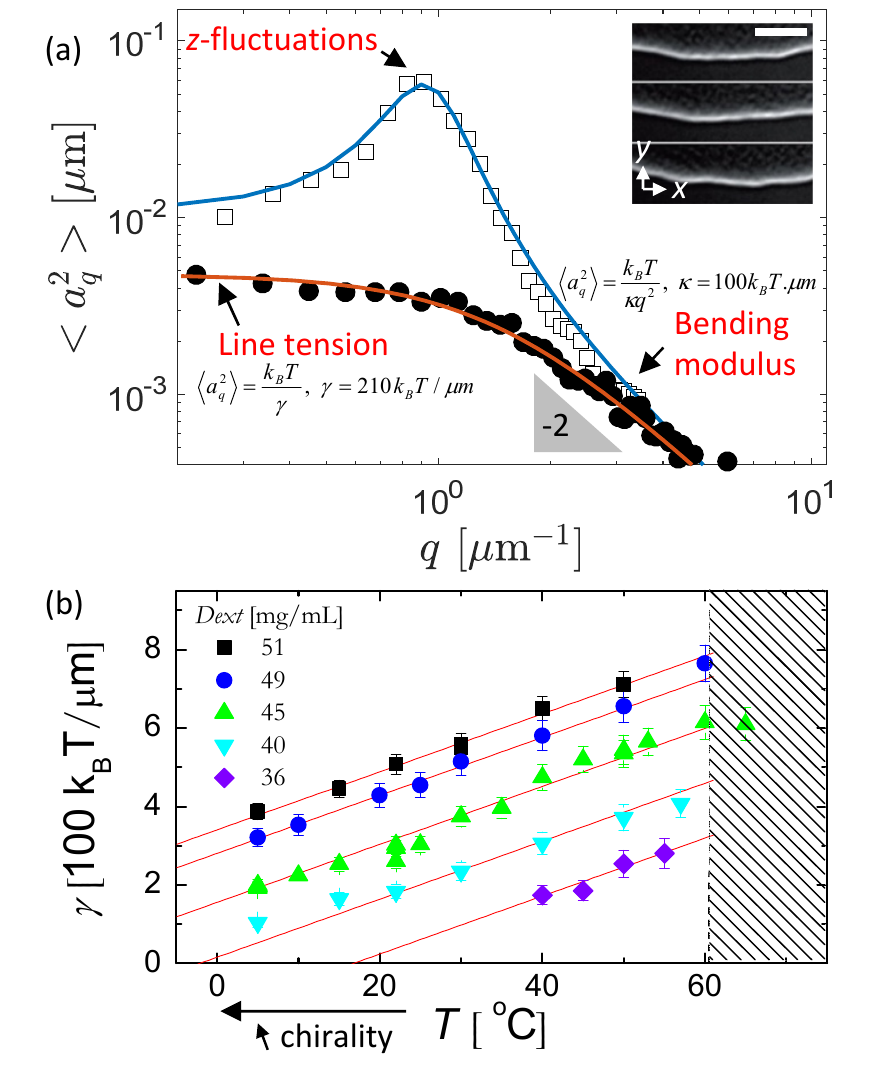}
     \caption{Chiral control of interfacial tension of colloidal membranes composed of fd-wt \cite{gibaud2012,jia2016,kang2016}. (a) Fluctuation spectrum of the membrane edge $\langle a_q^2 \rangle$ at $T=$ 20$^\circ$C (square) and 50$^\circ$C (circle) for $Dext=$ 36 mg/mL. For small wave number $q$, $\langle a_q^2 \rangle$ is independent of $q$ and inversely proportional to the effective line tension $\gamma$. For large $q$, $\langle a_q^2 \rangle$ are independent of $Dext$ and $T$ and scale as $1/q^2$. For very chiral samples a peak is observed at intermediate $q$ which is attributed to Gaussian curvature which enable out of plane fluctuations. Red lines are fits using equation (3). Inset: DIC images, taken 1s apart, illustrating the fluctuations of the membrane’s edge. Scale bar, 5$\mu$m. (b) Line tension as a function of temperature for different dextran concentrations. In the achiral limit at 60$^\circ$C, $\gamma_{chiral}$ = 0 and $\gamma_{bare}$ = $\gamma$. Increasing the Dextran concentration increases $\gamma_{bare}$. Decreasing the temperature reduces $\gamma$ because $\gamma_{chiral}$ decreases. The red lines of fixed slope are guides to the eyes illustrating the universal scaling of $\gamma$ with chirality.
     }
    \label{fig:lt}
\end{figure}

A classical way to measure the interfacial tension consists in analyzing the membrane’s edge thermal fluctuations in the Fourier space \cite{fradin2000,safran1994,aarts2004}. A typical fluctuation spectrum for an achiral edge is shown in Fig. \ref{fig:lt}. In the thermodynamic limit which corresponds to small wave vectors, $q$, the mean square Fourier amplitudes of the edge fluctuations, $\langle a_q^2 \rangle$, is $q$-independent, and yields the effective line tension, $\langle a_q^2 \rangle = k_BT/\gamma$ \cite{fradin2000}. In the large-$q$ limit, fluctuations scale as $1/q^2$ and yield the bending rigidity of the interface, $\kappa$. In the range of temperatures and dextran concentrations explored, for fd-wt colloidal membranes, $\kappa\sim$ 100 $k_BT.\mu m$ while $\gamma$ varies from $\sim$ 100 to 800 $k_BT/\mu m$

\begin{figure}
	\centering
  \includegraphics[width=8.5cm]{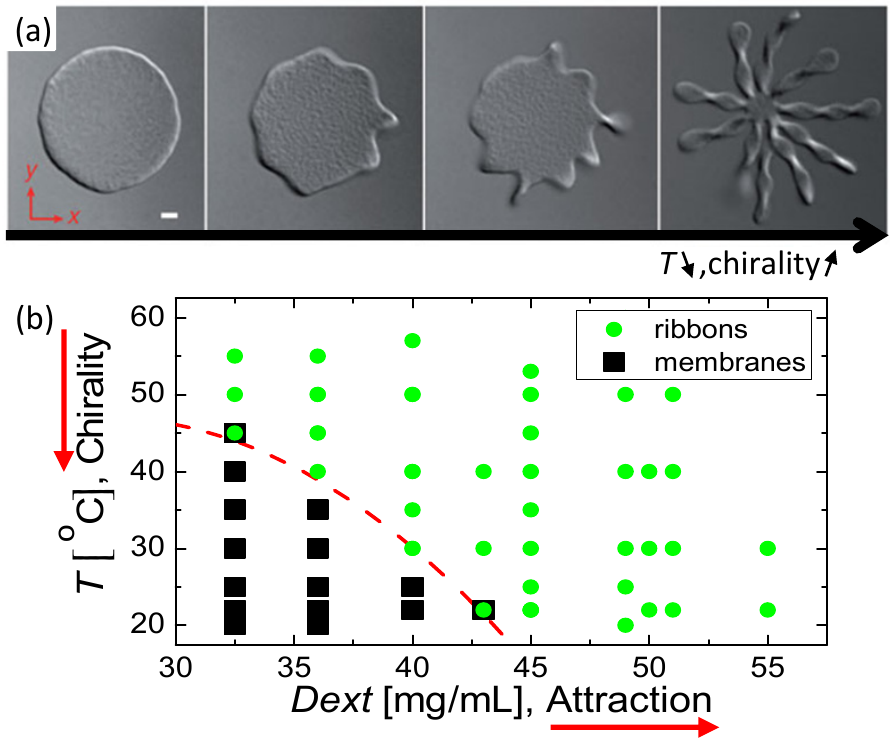}
     \caption{Reconfigurable self-assembly through chiral control of interfacial tension -- colloidal membrane to twisted ribbons transition, \cite{gibaud2012}. (a) DIC micrographes of a temperature quench  which show the transition of the 2D membrane into twisted ribbons. Scale bar, 2 $\mu$m. (b) The stability diagram indicates regions of phase space where twisted ribbons and membranes are observed as a function of $T$ and $Dext$. The dash boundary is a guide to the eye.
     }
    \label{fig:sun}
\end{figure}

Next we evidence the role of chirality on $\gamma$. The self-assembly of colloidal membranes is driven by entropy alone and therefore athermal as apposed to the fd-wt chiral interaction which depends solely on temperature. We thus have a unique system where it is possible to decorelate the effect of attraction (dextran concentration) from chirality (temperature): $\gamma(Dext, T)$ = $\gamma_{bare}(Dext) - \gamma_{chiral}(T)$, where $\gamma_{bare}$ is the bare line tension of a membrane edge composed of achiral rods and $\gamma_{chiral}$ is the chiral contribution to the line tension \cite{gibaud2012}. In Fig. \ref{fig:lt}, we observe that the effect of chirality drastically modifies the fluctuation spectrum of the edge of a colloidal membrane, $\langle a_q^2 \rangle$ as expected from the edge structure. First, $\langle a_q^2 \rangle$ is shifted upward at low $q$ which indicates that the line tension decreases with chirality as hypothesize. This is further demonstrated using dextran series which show that $\gamma$ decreases with the same slope as temperature decreases confirming that the two contributions to $\gamma$ are uncorrelated. Second, a peak appears at intermediate $q$. This peak is attributed to out of plane fluctuations. Indeed the effect of chirality at the edge of colloidal membranes is twofold. 2-D layered geometry cannot support twist and chirality is consequently expelled to the edges in a manner analogous to the expulsion of a magnetic field from superconductors \cite{de1972, renn1988}. Moreover, to palliate the 2D frustration \cite{kamien2001,renn1988, hough2009}, chirality forces the edge fluctuations to escape in the $z$-direction \cite{helfrich1988} which vouch for the existence of a positive Gaussian curvature, $\bar{k}\sim 150 k_BT$ \cite{jia2016}.

\subsubsection{Twisted ribbons}

\begin{figure}
	\centering
  \includegraphics[width=8.5cm]{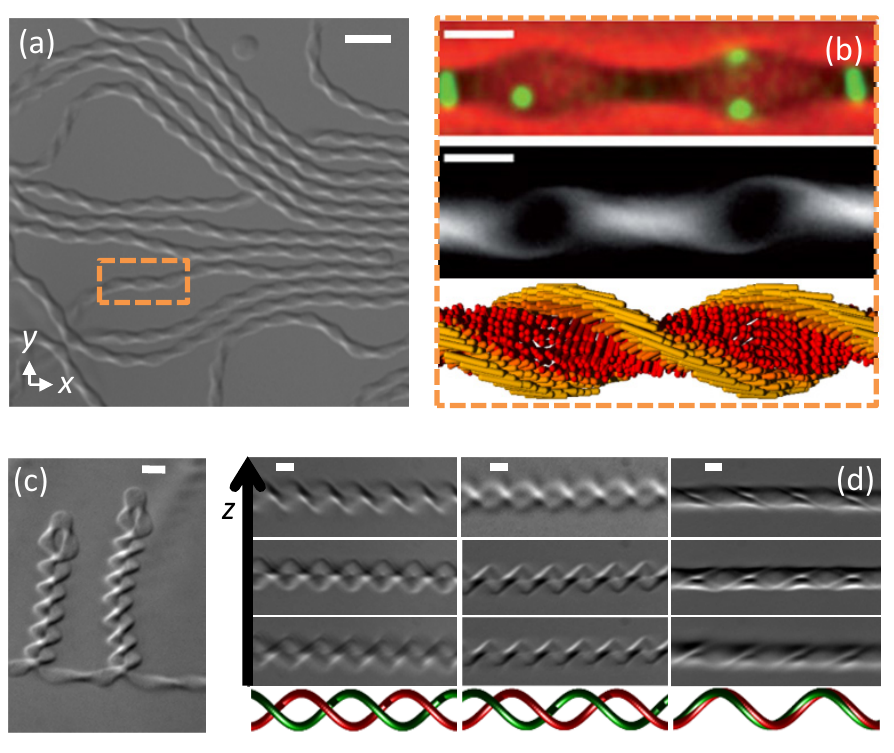}
     \caption{fd-wt twisted ribbons, $T=$ 22$^{\circ}$C, $Dext=$ 36 mg/mL.\cite{gibaud2012}. (a) DIC micrograph of twisted ribbons Scale bar 5 $\mu$m. (b) From top to bottom. An overlaid phase contrast (red) and fluorescence (green) image of a stable ribbon containing a low volume fraction of fluorescently labeled rods. LC-PolScope image indicating rod tilting that penetrates from the edge of the ribbon towards its center. Schematic structure of twisted ribbons as deduced from optical microscopy. (c) Doubly twisted ribbons consist of two twisted ribbons wrapped around each other. (d) $z$-stack DIC micrographs of doubly twisted ribbons with three different conformation. The red and green sinusoids follow the outer edges of each of the two ribbons comprising the double twisted ribbon, respectively and illustrate the phase shift between the two twisted ribbons can vary from being out of phase to being in phase. Scale bar 2 $\mu$m.
     }
    \label{fig:rib}
\end{figure}
 
The chiral control of line tension raises the possibility that at sufficiently low temperatures the chiral contribution to interfacial energy could dominate the bare line tension, lowering the energetic cost of creating edges and leading \th{to the control of the size of the membrane or to spontaneous edge formation. With decreasing temperature, membranes remain polydisperse in size. However} the membrane edge eventually becomes unstable, resulting in a remarkable polymorphic transition, Fig. \ref{fig:sun}. Twisted ribbons grow along the entire periphery of the disk from the out of plane fluctuations of the membrane edge, generating a starfish-shaped membrane. This polymorphic transition is reversible and twisted ribbons form equilibrium structures at high chirality and low dextran concentrations. Twisted ribbons are a beautiful example of hierarchical assembly, Fig. \ref{fig:rib}. It consists in a twisted monolayer of aligned rods which form a helicoidal structure perpendicular to the rod twist. As observed by Efrati and Irvine, such object is simultaneously right and left handed \cite{efrati2014}. The rod twist at the edge is left handed while on larger length scales the  helicoidal structure of the ribbon is right handed. As such it differs from other twisted ribbons observed in the literature \cite{zastavker1999, matsumoto2009, marini2002, zhang2002, srivastava2010, lashuel2000}. The twisted ribbons can be seen on a coarse grain level as polymers with a persistence length of the order of the pitch of the helicoidal structure and may form a zoology of structures ranging from branched polymer, to loop polymer or entangled phone cord like structures reminiscent of DNA doubled helix \cite{gibaud2012}. Twisted ribbons stability with respect to colloidal membranes is \th{attributed to two factors. First chirality is frustrated in colloidal membranes as viruses in the bulk cannot twist due to their virus neighbors whereas in twisted ribbons all the viruses twist and chirality may naturally be expressed. Second, twisted ribbons are edge objects with low interfacial energy compared to the membranes}. Given the 3D structure of the ribbons, Gaussian curvature may also be an important parameter that justify the twisted ribbon stability \cite{jia2016}.

\begin{figure}
	\centering
  \includegraphics[width=8.5cm]{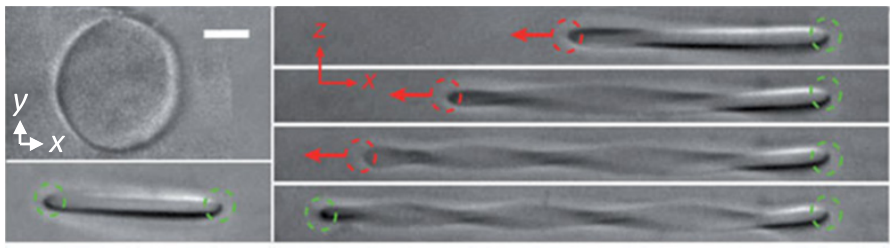}
     \caption{fd-wt colloidal membrane manipulation with optical tweezers, $T=$ 22$^{\circ}$C, $Dext =$ 45 mg/mL \cite{gibaud2012}. A colloidal membrane is trapped with a dual-beam optical trap and stretched, inducing the transition to a twisted ribbon. Red circles indicate moving traps and green circles indicate fixed traps. Bright field microscopy images. Scale bar, 2$\mu$m. 
     }
    \label{fig:tweez}
\end{figure}

It is possible to use laser tweezers to manipulate the self assembled structures. For instance in Fig. \ref{fig:tweez}, the two opposite sides of a colloidal membrane are trapped with a dual-beam optical trap. The viruses align with the electric field of the laser and the membrane turn sideway. Using a static trap and providing an extensional displacement with the other optical trap, the membrane is stretched, causing the transition  to a twisted ribbon. This mechanically induced disk-to-ribbon transition is reversible; on removal of the optical trap, the highly elastic ribbon relaxes back into its original shape. This experiment pave the way to study mechanical properties of self-assembled objects \cite{zakhary2014}. 

\subsection{Colloidal membranes and chiral coalescence}

Driven by the balance between interfacial tension and bulk energy, a pair of liquid droplets, when sufficiently close to one another, may coalesce to form a single daughter droplet. The coalescence process is complex and involve the rupture and the fusion of the droplets surfaces associated with energy barrier and local rearrangements \cite{aarts2005, aarts2008, paulsen2012, paulsen2011, sundararaj1995, loudet2000, chernomordik2008, haluska2006, jahn1999}. In most cases, it is an ‘all-or-none’ process; once initiated, the reaction proceeds to completion. However, there is also the possibility of incomplete coalescence. For example, vesicles coalesce into hemi-fused state \cite{shillcock2005} and nanotubes into a defect-ridden structure\cite{terrones2000}. Taking advantage of the chiral edge, colloidal membranes coalescence enlight the role of geometrical frustrations \cite{goodby1994} in the self-assembly of new structures \cite{zakhary2014b, gibaud2017}. 

\begin{figure}
	\centering
  \includegraphics[width=8.5cm]{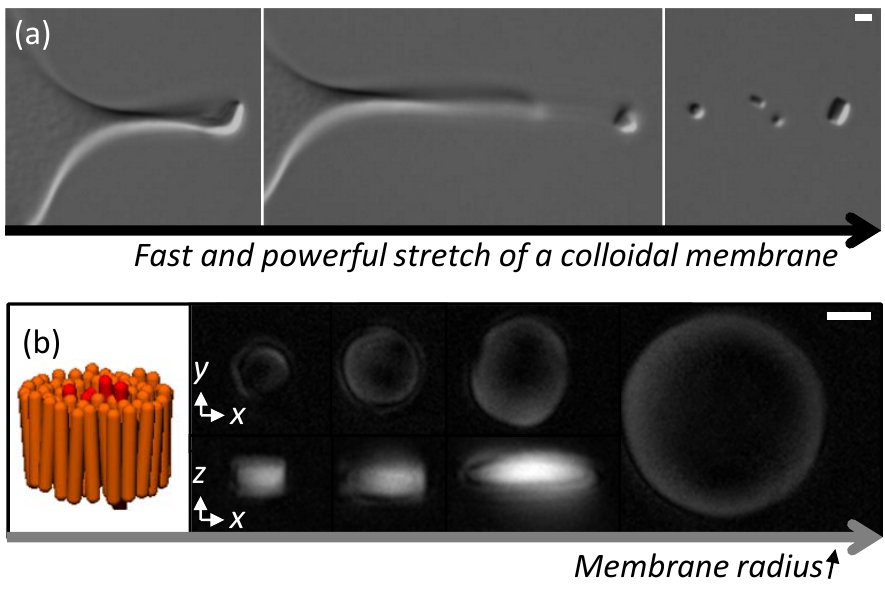}
     \caption{fd-wt membrane seeds, $T=$ 22$^{\circ}$C, $Dext =$ 45 mg/mL \cite{kang2016}. (a) DIC micrographs time sequence: using 2 Watt optical tweezers, it is possible to grab a colloidal membrane and detach membrane seeds. (b) Schematic of a membrane seed. 2D-LC-Polscope of membrane seed observed from the top and sideway. Scale bars 1 $\mu$m.
     }
    \label{fig:seed}
\end{figure}

\begin{figure}
	\centering
  \includegraphics[width=8.5cm]{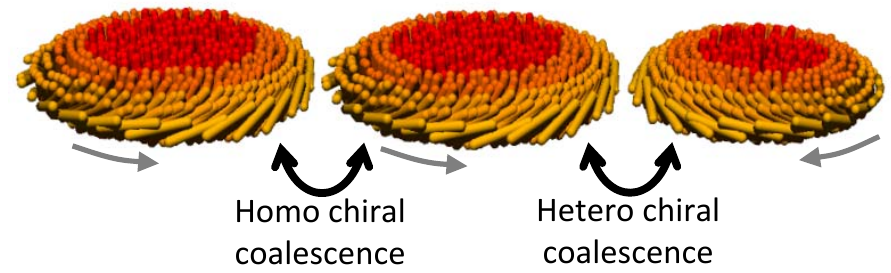}
     \caption{Chiral coalescence of colloidal membranes.
     }
    \label{fig:coal}
\end{figure}

The edge chirality of colloidal membranes can be controlled in various ways. First, if the colloidal membrane is small enough, a diameter smaller than the virus length, the viruses stand straight at the edge and those membrane seeds are achiral, Fig. \ref{fig:seed}. Second, for larger membranes the edge adopt the chirality of the virus. Third, it is possible to compose achiral virus suspensions, for instance using fd-wt at $T=$ 60$^\circ$C or mixtures of fd-wt and fd-y21m at $x_{fd}=$ 0.26, Fig.~\ref{fig:chol}-\ref{fig:chol2}, and in this case, the symmetry being broken, colloidal membranes self assemble either with left- or right-handed edge. For chiral membranes, we can divide the coalescence in two families: homo and hetero chiral coalescence, Fig. \ref{fig:coal}. In homo chiral coalescence two coplanar  membranes have the same chirality. The rods at the coalescence point are tilted in opposite directions, trapping 180$^\circ$ of twist between the two membrane. In hetero chiral coalescence both coplanar membranes have the opposite chirality and viruses at the edge need only to straiten in the $z$-direction at the coalescence point.  

\subsubsection{Homo chiral coalescence -- $\pi$-walls, pores, Möbius anchors and colloidal skyrmions}

\begin{figure}
	\centering
  \includegraphics[width=8.5cm]{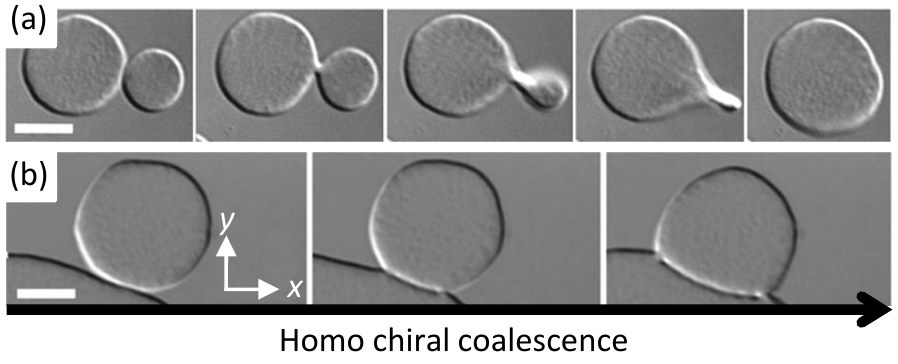}
     \caption{Homo chiral coalescence of fd-wt colloidal membranes, $Dext=$ 45 mg/mL, $T=$ 22$^{\circ}$C \cite{gibaud2017}. (a) DIC micrographs time sequence showing a coalescence pathway resulting in a defect-free daughter membrane. The coalescing membranes initially form a twisted bridge that induces a torque and a rotation by 180$^\circ$ that expel the twist between the two membranes. (b) DIC micrographs time sequence showing another coalescence pathway resulting in the formation of a $\pi$-wall. Because the membranes lie on the bottom surface, the rotation required for defect-free coalescence is suppressed. Scale bars, 10 $\mu$m.
     }
    \label{fig:homo}
\end{figure}

Homo chiral coalescence \cite{zakhary2014} is at the center of chiral topological frustration. In Fig. \ref{fig:homo}, the coalescence between two homo chiral membrane may result in a defect-free daughter membrane. Similar to liquid droplets, the thermal fluctuations are sufficient to form a bridge between the two membranes. In this one rod-length-wide bridge, the rods twist by 180$^\circ$ to match the orientations of the joining edges. The twisted bridge induces a torque which enable the two membranes to rotate along the axis formed by the bridge and expel the trapped twist. As the membranes twist around each other, the connecting bridge expands in width, eventually leading to a circularly shaped defect-free daughter membrane. In an other pathway, coalescence is initiated by the formation of two twisted anchors, which bind the membranes together and initiate the nucleation of a continuous 1D line defect. This line defect, named a $\pi$-wall, quickly grows to its equilibrium size, pushing the two anchor apart. Once the $\pi$-wall is fully formed, it remains indefinitely. $\pi$-walls are stable with respect to two free membrane edges: the measurements of $\pi$-wall interfacial tension $\gamma_{\pi}$ and the membrane edge interfacial tension $\gamma$ shows that $\gamma < \gamma_{\pi} < 2 \gamma$. 

\begin{figure}
	\centering
  \includegraphics[width=8.5cm]{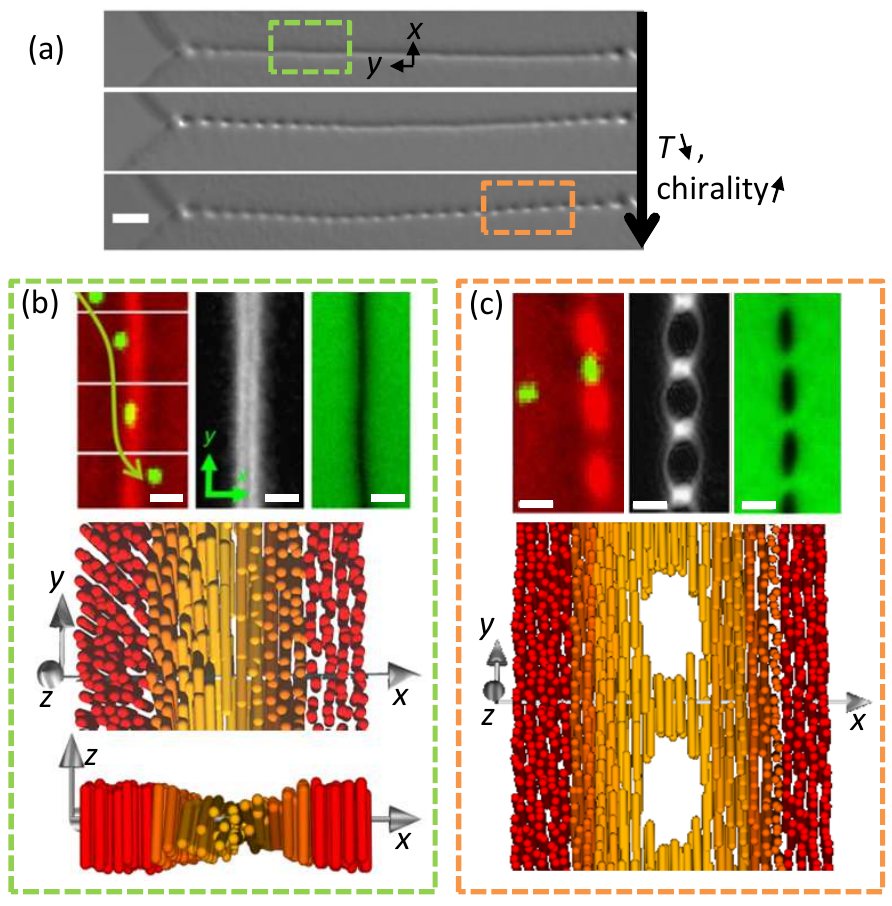}
     \caption{Reconfigurable self-assembly and chirality -- $\pi$-wall to alternating bridge-pore arrays (ABPA) transition in fd-wt colloidal membranes \cite{gibaud2017}. (a) A temperature quench induces a transition from a $\pi$-wall to an ABPA. The colloidal membrane assembled at $Dext=$ 45 mg/mL is quenched from $T=$ 55 to 22$^{\circ}$C. During the quench, the $\pi$-wall morph into alternating bridge-pore arrays. Similarly to the membrane/twisted ribbon transiton, this transition is reversible. DIC microscopy images. Scale bar, 5 $\mu$m. (b) $\pi$-wall structure. From left to right - A time sequence illustrates rotation of a fluorescently labeled virus (green) by 180$^\circ$ as it diffuses through a $\pi$-wall. Fluorescence images are supperposed with simultaneously acquired phase-contrast images (red). The 2D-LC-PolScope image of a $\pi$-wall is compatible with the 180$^\circ$ twist of the viruses. A fluorescence image of a $\pi$-wall where all the rods are fluorescently labeled. The defect appears darker in its center which correspond to a decreasse of the defect thickness. Bottom: sketch of $\pi$-wall. (c) ABPA structure. From left to right - The simultaneous fluorescence (green) and phase-contrast (red) imaging reveals that the rod twist by 180$^\circ$ whitin a bridge. A 2D-LC-PolScope confirms this twist. An ABPA image where all the rods are fluorescently labeled shows that the pores are empty of viruses. Bottom: sketch of ABPA. Scale bars, 2 $\mu$m.
     }
    \label{fig:pore}
\end{figure}
 
However, in fd-wt systems where chirality can be controlled with temperature, $\pi$-wall can be continuously brought to regimes with high chiral interactions at low temperatures. In this case, $\gamma_{\pi}>2\gamma$ and $\pi$-walls become metastable with respect to isolated membranes. In high chiral regimes, we do not observe the spontaneous dissociation of a $\pi$-wall into two defect-free membranes. We instead observe the opening of pores in the $\pi$-walls, Fig. \ref{fig:pore}. Those pores may form an alternating bridge-pore arrays (ABPA) structure which can be closed back into a $\pi$-wall by increasing the temperature. This behavior remains to be understood but seems reasonable, as pores create a large amount of edge interfaces which are favored at high chirality. More importantly, it empirically proves that with the proper ingredient it possible to actuate pores upon external signaling in self-assembled membranes. Pore actuation is of primal importance through out the cell life cycle \cite{suntharalingam2003, tran2006}.

Finally we discuss two structures related to $\pi$-walls: Möbius anchors and colloidal skyrmions. Both those structures rely on a robust on-demand method for imprinting defects into colloidal membranes with arbitrary spatial precision. Taking inspiration from recent work with thermotropic liquid crystals \cite{ackerman2012,honglawan2013,yoon2007}, we also use an optical trap. 

\begin{figure}
	\centering
  \includegraphics[width=8.5cm]{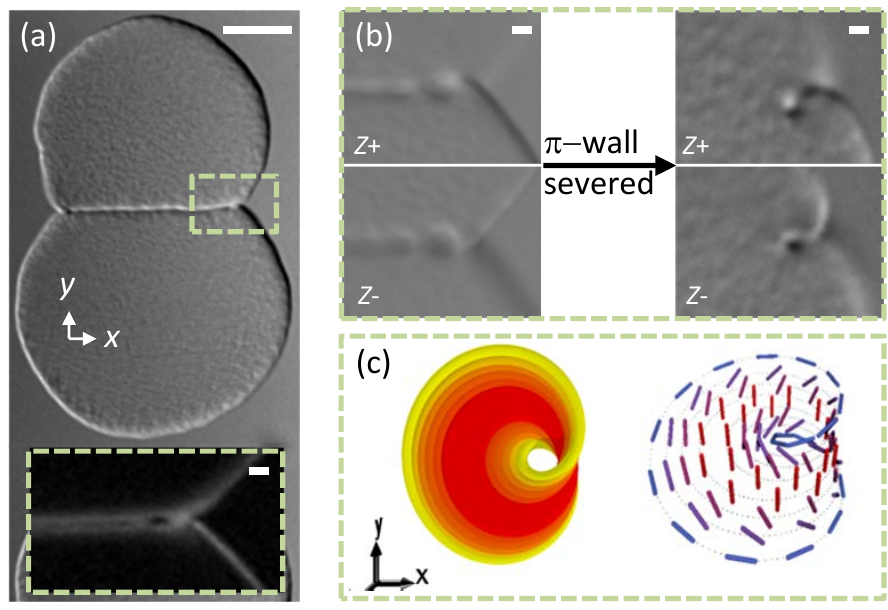}
     \caption{Möbius anchors in fd-wt colloidal membranes, $Dext=$ 45 mg/mL, $T=$ 22$^{\circ}$C \cite{zakhary2014}. (a) DIC micrograph of a $\pi$-wall, scale bar 5$\mu$m . Inset: 2D-LC-PolScope anchor point of the $\pi$-wall at the egde of a colloid membrane. We guess that this anchoring point forms a Möbius loop which prevents the $\pi$-wall to vanish. (b) DIC image of the Möbius anchors before and after the $\pi$-wall is severed with optical tweezers. $z+$ and $z-$ correspond to images focused slightly above and bellow the membrane plane. (c) Sketch of the Möbius anchor. Scale bars 2 $\mu$m.
     }
    \label{fig:mobius}
\end{figure}

A simple Möbius strip is a one-sided continuous surface, formed by twisting a long narrow rectangular strip of material through 180$^\circ$ and joining its ends. Such a structure can be made in liquid-crystal by knotting of microscopic topological defect lines with optical tweezers about colloids \cite{tkalec2011}. The Möbius strip we observe in colloidal membrane are Möbius anchors \cite{zakhary2014b}. The Möbius anchor is associated with the way $\pi$-walls are anchored to the membrane edge, Fig. \ref{fig:mobius} and is mandatory for the $\pi$-wall to remain stable. For instance, it is possible with optical tweezers to imprint $\pi$-wall on a colloidal membrane. However, if the optical trap is released before the $\pi$-wall is anchored, the defect retracts. Based on 2D-LC-Polscope micrographs, it seems that the viruses follow a simple Möbius strip which tight the $\pi$-wall to both edges of the daughter membranes, Fig. \ref{fig:mobius}. At this point this anchoring structure is only a guess. This hypothesis is however supported by the fact that it is necessary to produce a the back and forth motion with the optical trap to create the anchor which is reminiscent of the pathway depicted in soap film to create Möbius loop \cite{goldstein2010}.

\begin{figure}
	\centering
  \includegraphics[width=8.5cm]{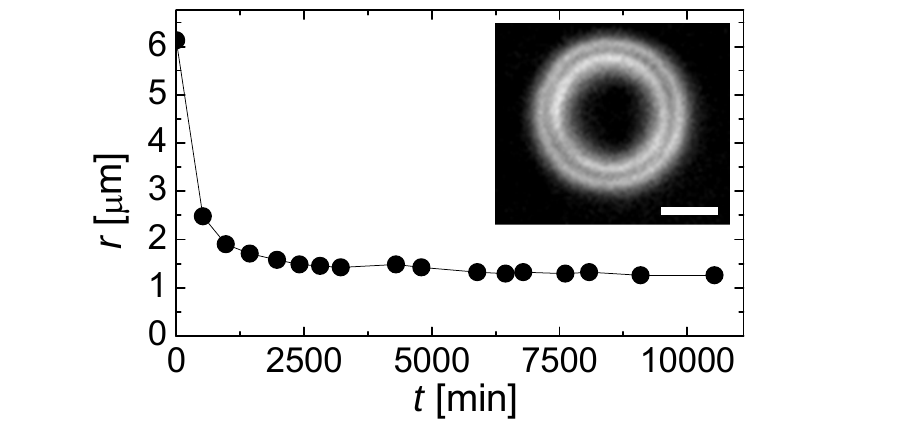}
     \caption{Colloidal skyrmion in fd-wt colloidal membranes, $Dext=$ 45 mg/mL, $T=$ 22$^{\circ}$C \cite{zakhary2014}. Evolution of the radius of the colloidal skyrmion over time. $t=0$ correspond to the time the colloidal skyrmion is imprinted with optical tweezers. Inset: 2D-LC-PolScope image of a colloidal Skyrmion using laser tweezers to close a $\pi$-wall on itself. Scale bar 1 $\mu$m.
     }
    \label{fig:skyr}
\end{figure}

Colloidal skyrmions \cite{zakhary2014b} are obtained using optical tweezers to cleave a $\pi$-wall in two places, and then quickly joining the two ends to form a closed ring embedded within the membrane, Fig. \ref{fig:skyr}. The colloidal skyrmion shrinks to an equilibrium diameter of about $\sim$ 1 $\mu$m. Note that for a similar size, an isolated colloidal membrane displays an untwist edge, Fig. \ref{fig:seed}. The colloidal skyrmions share properties with skyrmion excitations encountered in hard condensed matter physics \cite{fukuda2011, muhlbauer2009, brey1995, tonomura2012, ravnik2011}. It is topologically protected \cite{skyrme1962,nagaosa2013}: it has a positive energy compared with the background field but the $\pi$-wall forming the skyrmion cannot be untrapped unless the $\pi$-wall is severed. Moreover, it is a 2D structure characterized by a vorticity  $m=1$ and a phase helicity $\varphi=\pm \pi/2$ which sign depends on the chirality of the $\pi$-wall. As such, it is very similar to singled out skyrmions from the hexagonal SkX state on MnSi \cite{muhlbauer2009} and Fe$_{1 - x}$Co$_x$Si \cite{grigoriev2007, grigoriev2009, yu2010} and seems to be the closest realization of a theoretical nematic skyrmions restricted to straight infinite lines in unbounded ideal materials \cite{bogdanov2003}. It however differs from other liquid crystal skyrmions such as double twist cylinders “baby-skyrmions” \cite{ackerman2014, smalyukh2010}, skyrmions in cholesteric blue phases subjected to strong external fields \cite{hornreich1989, heppke1989}.

Understanding the principles that support or prevent membrane coarsening and defects formation such as $\pi$-wall is essential to grow large defect free membranes and consider applications. As chirality is at the center of $\pi$-walls, it suggests that producing achiral colloidal membranes would lead to defect-free coalescence and uniform monolayers.

\subsubsection{Hetero chiral coalescence -- scalloped membranes and gaussian curvature} 

\begin{figure}
	\centering
  \includegraphics[width=8.5cm]{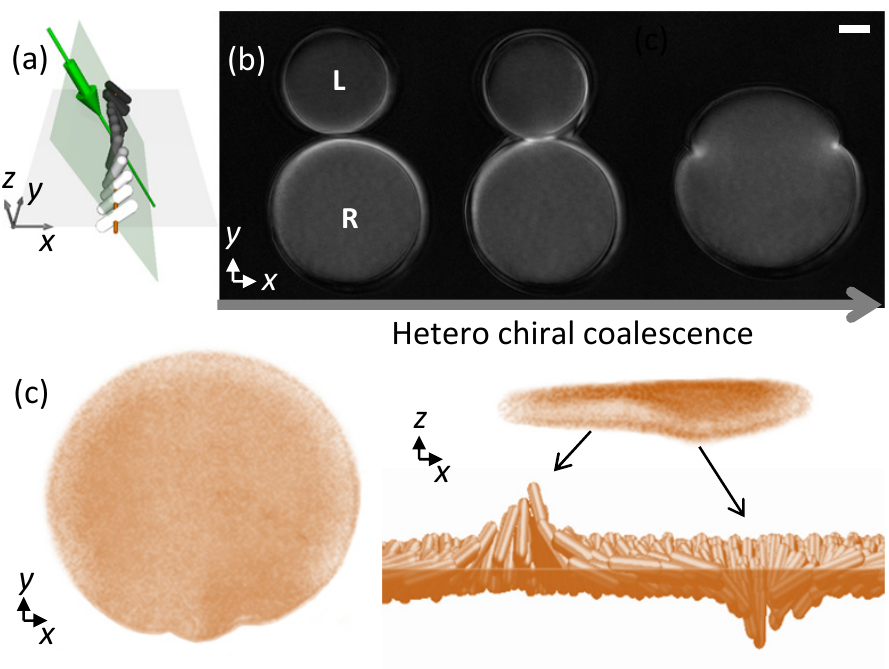}
     \caption{Hetero chiral coalescence leads to scalloped membranes \cite{gibaud2017}, in mixtures of fd-wt and fd-y21m, $x_{fd}=$ 0.26, $Dext=$ 45mg/mL, $T=$ 22$^{\circ}$C. (a) Sketch of 2D-LC-Polscope Angled light illuminating a 1D cross-section of twisted rods. Because the direction of the incident light is tilted towards the x axis, rods twisting counter-clockwise along the light source are not birefringent and appear dark. In contrast, rods twisting clockwise, away from the incoming light, have higher optical anisotropy and thus appear bright. Tilting the light source breaks the symmetry of the 2D-LC-PolScope setup and allows us to distinguish between left-handed (L) and right-handed (R) membranes. The gray scale changes from dark to light with increasing retardance, where the rods aligned with the direction of incident light have zero retardance. (b) 2D-LC-Polscope Angled light micrographs time sequence showing the coalescence of a left-handed membrane with a right-handed membrane. The daugher membrane displays two cusps that separate the left-handed edge section from the right-handed one. Scale bars, 4 $\mu$m. (c) Reconstruction of the membrane based on the confocal images and sketch of the point defects. Two adjacent point defect escape in the third dimension $z$ with opposite direction. 
     }
    \label{fig:hetero}
\end{figure}

To study hetero chiral coalescence \cite{gibaud2017}, colloidal membranes composed of homogeneous mixtures of fd-wt and fd-y21m are used. For 0.04  $< x_{fd} <$ 0.45, in the early stages of the sample maturation, we observe colloidal membranes of either edge handedness, indicating spontaneously broken achiral symmetry. Over time, the intermediate-sized membranes with mixed edge twist continue to coalesce. Both homo and hetero chiral coalescence is observed. In both cases coalesced membranes display an homogeneous mixing of fd-wt and fd-y21m. In hetero chiral coalescence, as the two proximal edges of a pair of coplanar membranes merge, the twist of the edge-bound rods is expelled by aligning the constituent rods with the membrane normal. Hetero chiral coalescence leads to scalloped membranes. As compared to homo chiral coalescence, scalloped membranes form easily. Moreover, they are defect free in their bulk and may reach millimeter diameter. The hallmark of scalloped membranes is located on its edge. It displays two outward protrusions which separate a left from a right handed edge, Fig. \ref{fig:hetero}. Using confocal microscopy, it is observed that the two protrusions escape in the $z$-direction in opposite directions. This 3D point-like singularity on the vertical axis vouch for the presence of Gaussian curvature $\kappa_G$ associated with its Gaussian elastic modulus $\bar{k}$.

\begin{figure}
	\centering
  \includegraphics[width=8.5cm]{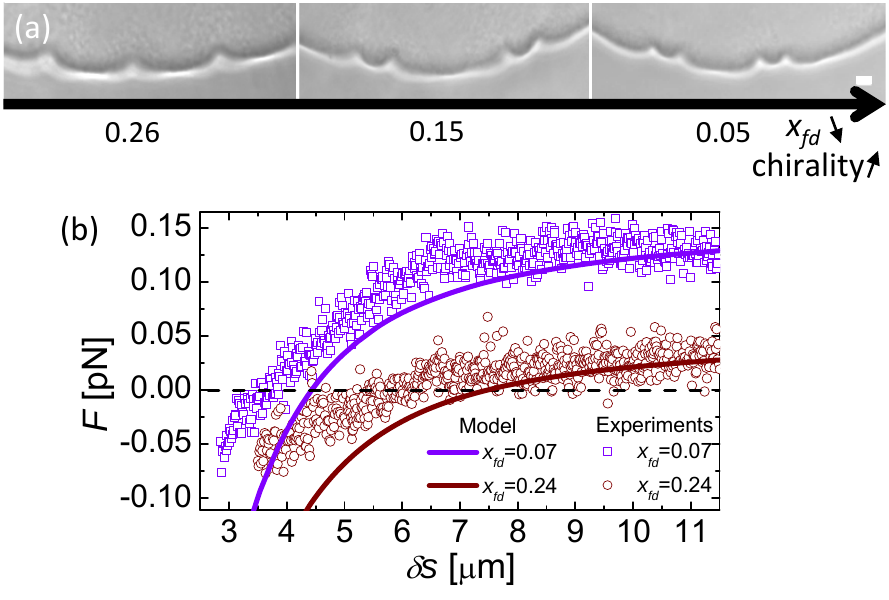}
     \caption{Measurement of the protrusion interactions in fd-wt/fd-y21m scalloped membranes, $T=$ 22$^{\circ}$C, $Dext =$ 45 mg/mL \cite{gibaud2017}. (a) Phase contrast image of scalloped membranes at $Dext=$40 mg/mL. Decreasing ratio of $x_{fd}$ (increasing the chirality) leads to a tighter coupling between two adjacent protrusion which then pair. Scale bars, 2 $\mu$m. (b) Force measurements $F$ (dots), obtained with laser tweezers, and fitted with a theoretical model (full curves) as a function of $\delta s$, the distance between two adjacent protrusions.
     }
    \label{fig:inter}
\end{figure}

The distance $\delta s$ between two adjacent edge protrusion greatly depends on $x_{fd}$, Fig. \ref{fig:inter}. Close to the achiral limit, at $x_{fd} =$ 0.26, adjacent protrusions freely move along the edge and the dynamics of $\delta s$ is diffusive. On the contrary, close to the boundary of the stability region of scalloped membranes at $x_{fd}=$ 0.04 or 0.45, adjacent protrusions pair and remain bound to each other at a well-defined distance $\delta s_0$. To measure the entire binding potential, active experiments are performed: one defect is moved by $\delta s$ using an optical trap, while simultaneously measuring the force $F$ exerted on the adjacent defect. For this purpose, 1.5 $\mu$m diameter colloidal beads are embedded into two adjoining cusp defects. The force is negative below $\delta s_0$, and positive above $\delta s_0$. $\delta s_0$ is the stable equilibrium position. The force steeply increases for small separations and saturates at large separations, indicating that a pair of defect is permanently bound, Fig. \ref{fig:inter}.

These observations can be mainly explained by the chirality of the edges. For instance, at $x_{fd}=$0.07 the system is mostly composed of fd-y21m and right handed chirality is favored. Therefore right-handed edge have a lower energy than left handed edges. This leads, in scalloped membranes, to a finite difference in line tension $\Delta \gamma$ between the left-handed and right-handed outward protrusions. The edge free energy is minimized by reducing the length of the outward protrusions with the unfavored twist and the amplitude of $\Delta \gamma$ accounts for the strength of the long range attraction between two adjacent protrusion. Approaching the two protrusion close together has two consequences. It tends to over bend the edge separating the two adjacent protrusion and to flatten the protrusion in the $z$-direction. This works again the bending rigidity of the edge $\kappa$ and against a negative Gaussian curvature $\kappa_G$, which lowers the free energy of elastic deformations if the Gaussian modulus is positive and sufficiently large, $\bar{k}=$200 $k_BT$ \cite{tu2013,kaplan2013, gibaud2017}. 

This results display a striking difference with conventional bilayers which have a negative Gaussian modulus: saddle-shaped deformations increase the membrane energy \cite{hu2012,siegel2004,semrau2008,baumgart2005}. Moreover, scalloped membranes and the transition from membrane to twisted ribbons demonstrate that simple uniform elastic sheets lacking in-plane rigidity can spontaneously assume complex 3D folding patterns as opposed to thin elastic sheets with in-plane elasticity \cite{cerda2003, leocmach2015, king2012, pocivavsek2008} which require in-plane heterogeneities or an external force to be fold or wrinkled. Finally achiral symmetry  breaking has  been  observed  in  diverse soft systems  with orientational  order, ranging from lipid monolayers and nematic tactoids to confined chromonic liquid crystals \cite{tortora2011,jeong2014, link1997,jeong2015,hough2009c}. In particular the measured structure  and  interactions  of  the  cusp-like defects in  colloidal  membranes  resemble  studies of point defects moving along a liquid crystalline dislocation line in the presence of chiral  additives \cite{zywocinski2005}. The  main  difference is that in the colloidal membranes the achiral symmetry breaking leads to out-of-plane 3D membrane distortions that  couples  liquid  crystal  physics to membrane deformations. This is not possible for inherently confined liquid crystalline films.

\subsection{Asymmetric mixtures of colloidal rods}

\subsubsection{Phase separation in colloidal membranes}

\begin{figure}
	\centering
  \includegraphics[width=8.5cm]{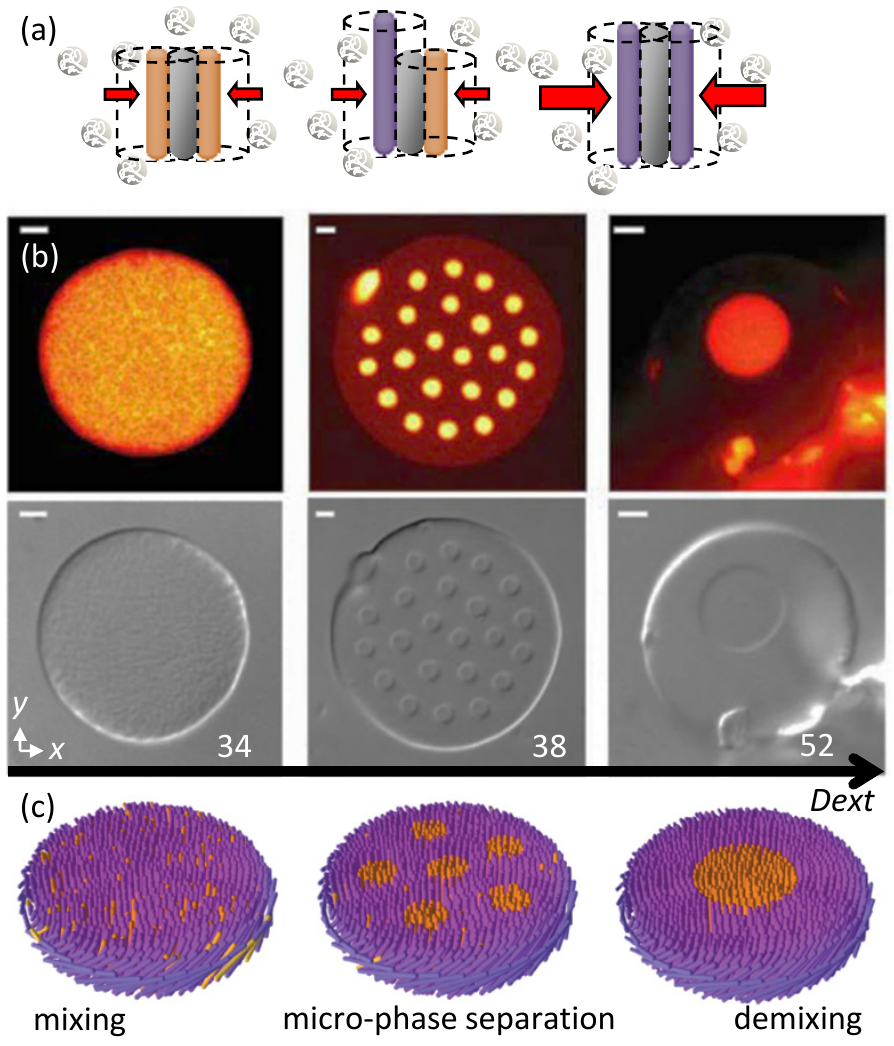}
     \caption{Phase separation and membrane rafts in mixture of viruses fd-y21m and MK13KO7 \cite{sharma2014, kang2017}. (a) Sketch of the depletion interaction in mixture of long and short rods. (b) Dual view Fluorescence images of a fd-y21m (yellow)–MK13KO7 (red) and DIC micrograph of colloidal membrane as a function of dextran concentration. At $Dext=$ 34mg/mL the membrane is homogeneously mixed. At $Dext=$ 38mg/mL, we observe the formation of finite-sized clusters enriched in fd-y21m embedded in a background enriched in MK13KO7. At $Dext=$ 52mg/mL, complete separation of the bulk phase is observed. Scale bar, 5$\mu$m. (c) Sketch of the colloidal membrane in (b).
     }
    \label{fig:ps}
\end{figure}

Phase separation can be triggered by asymmetric forces between the colloids. This force configuration can be achieve by mixing depletant with viruses of different lengths: fd-y21m virus (880 nm long) and M13KO7 virus (1200 nm long). The strength of the depletion force is proportional the overlap of the excluded volume. In fig. \ref{fig:ps}, two short rods and a short rod and long rod share the identical overlap of the excluded volume while two long rods have a large overlap of the excluded volume and therefore display greater attraction.

Colloidal membranes containing both fd-y21m (right handed) and M13Ko7 (left handed) are assembled by adding a depletant to a dilute isotropic mixture of fd-y21m and M13KO7, \cite{sharma2014}. After reaching a large enough size, membranes sediment to the bottom of the sample chambers; the constituent rods pointed in the $z$ direction, Fig. \ref{fig:ps}. At low depletant concentrations, thermal energy is sufficient to overcome the attraction between the rods of different sizes and the rods remain homogeneously mixed in the membrane. At high depletant concentrations the rod within the membrane separate into two phases: an enriched M13KO7 phase surround by an enriched fd-y21m phase. Both phases conserve the symmetry of the colloidal membrane. At intermediate concentrations, micro-phase separation is observed: colloidal rafts, highly monodisperse micrometre-sized 2D droplets enriched in fd-y21m, float in the background of M13KO7. 

\subsubsection{Membrane rafts}

\begin{figure}
	\centering
  \includegraphics[width=8.5cm]{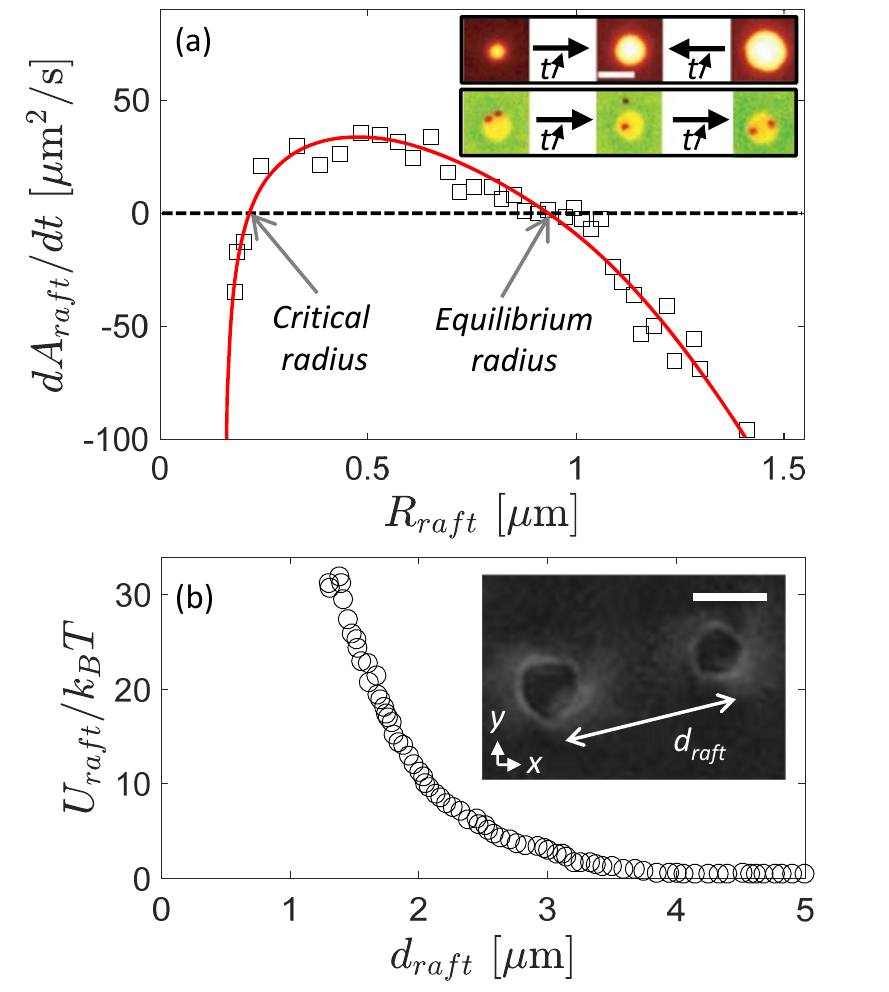}
     \caption{Equilibrium raft and their interaction in fd-y21m/M13KO7 colloidal membranes \cite{sharma2014}. (a) The dependence of raft expansion rates on raft size directly reveals the critical nucleus size and equilibrium size. Inset top: Expansion of an undersized raft and contraction of an oversized raft. Inset bottom: Fluorescently labeled fd-y21m rods associate and dissociate from a raft, revealing binding kinetics at the single-molecule level. (b) Effective pair interaction potential obtained using the blinking trap technique for clusters with diameters of 1.6 $\mu$m. Inset: 2D-LC-PolScope image of rafts embedded in a colloidal membrane. 
     }
    \label{fig:raft}
\end{figure}

Colloidal rafts \cite{sharma2014} do not coarsen with time, suggesting that they are equilibrium structures, Fig. \ref{fig:raft}. Particle  tracking  experiments  show  that the rods  diffuse in and out of these rafts, allowing for  equilibration to a preferred size. Using optical tweezers to create a raft population with heterogeneous radii, the raft growth rate is measured. Below a critical radius the rafts melt and above the rafts converge toward an equilibrium radius of $\sim$ 1 $\mu$m.

Colloidal rafts seem similar to equilibrium clusters found in protein and colloidal dispersions \cite{stradner2004, segre2001, groenewold2001}. The stability of equilibrium clusters is attributed to the mixed potential of the particles forming the cluster. This mixed potential is composed of a short range attraction and a long range repulsion. For the colloidal rods, the attraction is due to the depletion interaction. Contrary to equilibrium cluster particles, the electrostatic interactions of the colloidal rods are fully screened and the long range repulsion is attributed to virus chirality. Two raft are indeed in a homo chiral coalescence configuration which is not propitious for merging in 2D. The raft edge twist is further transmitted by the twisted structure of the background membrane which mediates a long range elastic repulsion between rafts. This interaction is measured quantitatively by bringing two rafts close together with optical traps and tracking their trajectories upon release of the traps \cite{sharma2014}. This  chiral repulsion  stabilizes  small  rafts  against  an interfacial  line  tension  that  would  otherwise  promote  coarsening to a single raft domain and establishes a preferred depletant-concentration–dependent raft size \cite{kang2017}.

Those results fuel the ongoing discussion on the lipid raft which structure, properties and function constitute ongoing research \cite{lingwood2010, simons2004, hancock2006, klemm2009}. These membrane raft structures have evolved from controversial detergent-resistant entities to dynamic, nanometer-sized membrane domains formed by sterols, sphingolipids, saturated glycerophospholipids, and proteins \cite{dietrich2001, veatch2002, baumgart2003, simons2004, hancock2006}. Provided that the analogy between colloidal raft and lipid raft hold, it seems that short range attraction and chirality are the essential ingredient. A systematic study of the role a chirality in colloidal rafts with respect to the chiral molecule present in lipid raft remains to be done for a more refined analogy.


\section{Conclusion and perspectives}
\label{sec:conclusion}
\begin{figure*}
	\centering
  \includegraphics[width=17cm]{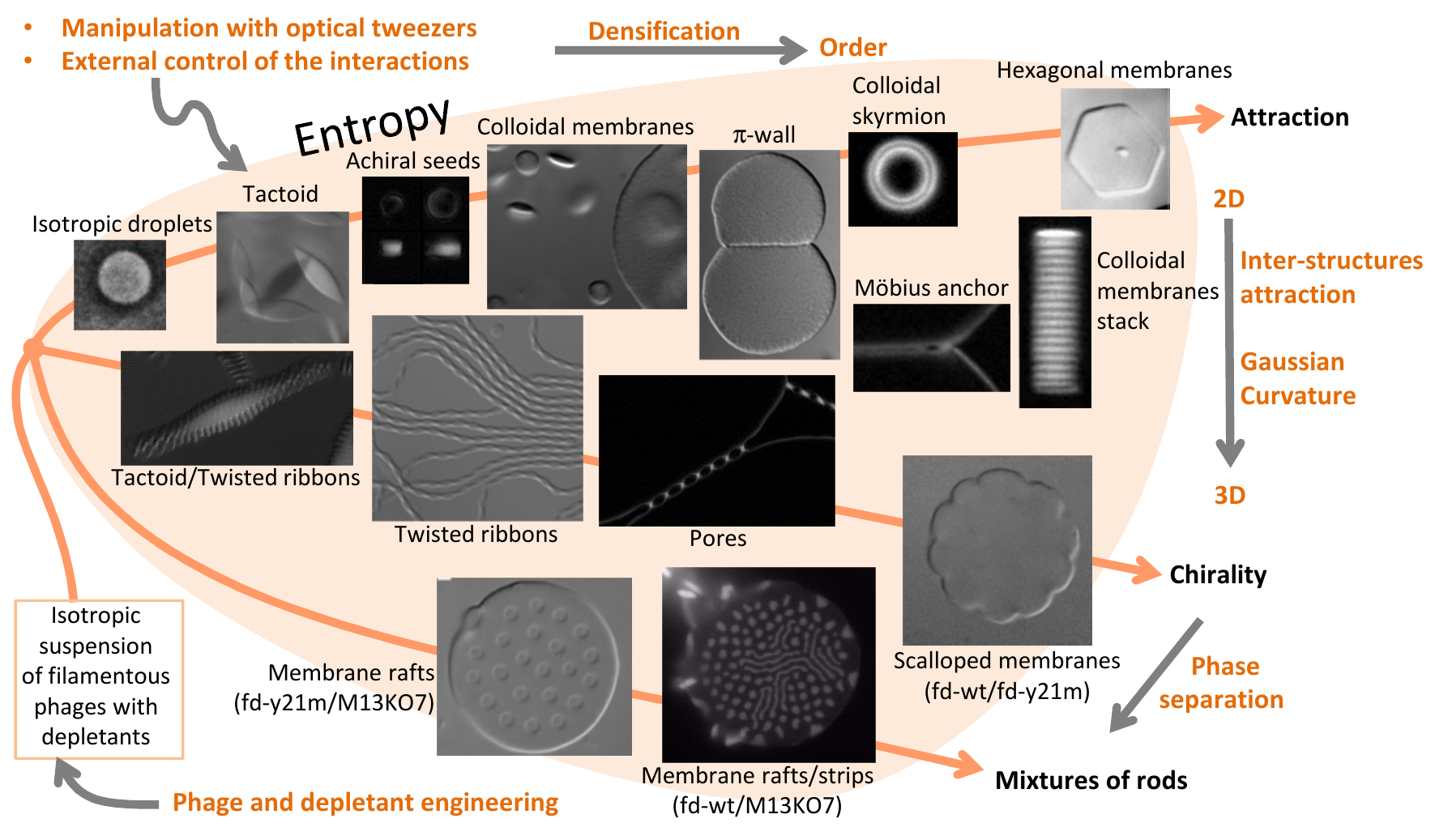}
     \caption{Filamentous phages as building block for reconfigurable and hierarchical self-assembly. Starting with isotropic suspensions of filamentous phages and depletion interactions, it is possible to drive the system to self assemble in a myriad of structures. Tuning the depletion interaction leads to condensate the virus with increasing density and order. Using the virus chirality, it is possible to turn 2D structures in 3D structures with Gaussian curvature. By mixing different type of rods the system may phase separate. Optical tweezers and external control over the interactions permits to navigate continuously in this state diagram and trigger shape-shifting structures.
     }
    \label{fig:fin}
\end{figure*}

Filamentous phage such as fd-like viruses are rod-like colloids that have well defined properties such as their diameter, length, rigidity, charge and chirality. Engineering those virus leads to a library of rods with slightly different properties which can be used as building blocks for self assembly, section \ref{sec:fil}. Their condensation in aqueous solution with additive depletants produces a myriad of structures ranging from isotropic/nematic droplets \cite{modlinska2015}, colloid membranes \cite{barry2010, yang2012,barry2009a}, achiral membrane seeds \cite{kang2016}, twisted ribbons \cite{gibaud2012}, $\pi$-wall \cite{zakhary2014}, pores, colloidal skyrmions, Möbius anchors, scallop membranes \cite{gibaud2017} to membrane rafts \cite{sharma2014}, section \ref{sec:con}. First, those structure reinforce the general notion that through a careful choice of particle shapes, sizes, and concentrations it is possible to “engineer entropy” \cite{frenkel2015} and build structures of ever-increasing complexity. Second, the entropy driven condensation of millions of rods in finite liquid-like objects leads to dynamic equilibrium and allows the structures to permanently rearrange and test their energy landscape. Therefore, those structures are very sensitive to externally tunable interactions like chirality and attractions which trigger shape shifting transitions. Third, external forces like optical tweezers may be utilize to manipulate those structures, probe their mechanical properties and the transition between multiple metastable polymorphic forms with complex topologies. Fourth, those structures represent a showcase of analogies between objects which belong to different fields of science such as colloidal membranes and lipid bilayers, chiral pore actuation and pores in cells, colloidal rafts and membrane rafts, colloidal skyrmions and solid state skyrmions, the twist penetration length at the edge of colloidal membranes and the penetration depth of the magnetic field in superconductor, or  Möbius anchors and Möbius strips. Fifth, this experiments work combined with theoretical inputs makes it well establish field in self-assembly. \th{Many theoretical approaches have been proposed. A de Gennes framework accompanied by appropriate surface energy terms was used to characterize colloidal membranes, twisted ribbons and $\pi$-wall \cite{tu2013, tu2013b, kaplan2013, kaplan2010, kaplan2014, yang2011, zakhary2014}. Sakhadande \textit{et al.} adopted a continuum Ginzburg-Landau theory to study raft stability \cite{sakhardande2016}. Xie \textit{et al.} considered a functional density theory constructed on the free volume theory for depletant-rod interactions, and a third order virial expansion for rod-rod interactions, with the equation of state for a hard disk system to constrain the areal rod density to study 2D colloidal membranes composed of binary mixture of rods with opposing chiralities \cite{xie2016}. Kang \textit{et al.} formulated an entropically-motivated theory using three simple considerations to characterize colloidal membranes and membrane rafts stability: depletant excluded volume, rod fluctuations perpendicular to the membrane, and rod twisting as described by the Frank free energy \cite{kang2017, kang2016}.} For all those reasons, fd-like phages constitute an attractive model system in soft matter physics, Fig.~\ref{fig:fin}.

The subject is clearly open and many questions remain unanswered. 2D colloidal membranes do not form vesicles -- would it possible with smaller viruses to reduce the lateral bending rigidity of the membranes and have them form vesicles? We have seen that chirality tends to produce 3D structures with gaussian curvature -- is possible to enhance this effect to make 3D leather pouch like membranes? Raft are stabilized due to chirality -- what happen to micro phase separation in homo chiral mixtures and in achiral mixutures? This review  being only based on three different phages (fd-wt, fd-y21m and M13KO7) which is really far from being representative of phage diversity \cite{rohwer2003}, there are many more structures to be discovered in such systems.

The phenomenology described in this review article should be relevant to diverse colloidal and nanosized rods that interact through excluded volume interactions. Indeed, as demonstrated in section \ref{sec:fil}, fd-like viruses are an excellent experimental realization of hard rods. The challenges for applications, especially in materials science, are threefold. Firstly, it lies in the development of monodisperse rods with interactions that exclude aggregation and permits equilibrium self assembly. Secondly, it necessitate robust rods that conserve their integrity in harsh conditions. Thirdly, large scale productions is required. Progress in these directions are clearly on their way \cite{koenderink1999, carbone2007, querner2008, kuijk2014, zhou2012} and material and bio-applications line up \cite{mao2009, jin2016, farr2014, yang2013}: templates for cells growth \cite{merzlyak2009}, colourimetric sensors \cite{oh2014}, photovoltaic devices \cite{dang2011, chiang2012}, batteries \cite{nam2006, lee2009, royston2008}, etc. \ldots

\section*{Acknowledgement}
I sincerely thank Zvonimir Dogic who introduces me to the subject of self-assembly and filamentous phages. Many thanks to Edward Barry, Anna Modlińska, Prerna Sharma, Andrew Ward and Mark J. Zakhary for countless hours spent together in the lab making those experiments work; to C. Nadir Kaplan, Louis Kang, Tom C. Lubensky, Robert B. Meyer, Robert A. Pelcovits, Thomas R. Powers and Hao Tu for their theoretical insight; and Seth Fraden, Eric Grelet, Pavlik Lettinga and Rudolf Oldenbourg for useful discussions.


\bibliography{biblio}

\end{document}